\documentclass[preprint1, 12pt]{aastex}
\shorttitle{Observations of VV 114}
\shortauthors{Sliwa et al.}

\newcommand{\tcoone}{$^{13}$CO $J$=1-0}
\newcommand{\coone}{$^{12}$CO $J$=1-0}
\newcommand{\cotwo}{$^{12}$CO $J$=2-1}
\newcommand{\tcotwo}{$^{13}$CO $J$=2-1}
\newcommand{\cothree}{$^{12}$CO $J$=3-2}
\newcommand{\cosix}{$^{12}$CO $J$=6-5}
\newcommand{\lfir}{$L_{\rm{FIR}}$}
\newcommand{\lsol}{$L_{\odot}$}
\newcommand{\msol}{$M_{\odot}$}

\newcommand{\kms}{km s$^{-1}$}

\newcommand{\tkin}{$T_{\rm{kin}}$} 
\newcommand{\nhtwo}{$n_{\rm{H_{2}}}$}
\newcommand{\alphaco}{$M_{\odot}$ (K km s$^{-1}$ pc$^{2}$)$^{-1}$}
\newcommand{\nco}{$N_{\rm{^{12}CO}}$}
\newcommand{\ff}{$\Phi_{\rm{A}}$}
\newcommand{\nmol}{$N_{\rm{mol}}$}
\newcommand{\xco}{[$^{12}$CO]/[$^{13}$CO]}
\newcommand{\nthreeco}{$N_{\rm{^{13}CO}}$}
\newcommand{\xh}{$x_{\rm{CO}}$}

\begin{document}
\title{Luminous Infrared Galaxies With the Submillimeter Array. \\ IV. \cosix\  Observations of  VV 114}

\author{Kazimierz Sliwa\altaffilmark{1},  
	Christine D. Wilson\altaffilmark{1},
	Melanie Krips\altaffilmark{2}, 
	Glen R. Petitpas\altaffilmark{3}
	Daisuke Iono\altaffilmark{4},
	Mika Juvela\altaffilmark{5}
	Satoki Matsushita\altaffilmark{6},
	Alison Peck\altaffilmark{7} and
	Min Yun\altaffilmark{8}}
 
\altaffiltext{1}{Department of Physics and Astronomy, McMaster University, Hamilton, ON L8S 4M1 Canada; sliwak@mcmaster.ca, wilson@physics.mcmaster.ca}
\altaffiltext{2}{Institut de Radio Astronomie Millimetrique, 300 Rue de la Piscine, Domaine Universitaire, F-38406 Saint Martin dÕHeres, France;  krips@iram.fr}
\altaffiltext{3}{Harvard-Smithsonian Center for Astrophysics, Cambridge, MA 02138; gpetitpa@cfa.harvard.edu}
\altaffiltext{4}{Chile Observatory, National Astronomical Observatory of Japan, 2-21-1, Osawa, Mitaka, Tokyo 181-8588 d.iono@nao.ac.jp}
\altaffiltext{5}{University of Helsinki, Finland; mika.juvela@helsinki.fi}
\altaffiltext{6}{Academia Sinica Institute of Astronomy and Astrophysics, P.O. Box 23-141, Taipei 10617, Taiwan; satoki@asiaa.sinica.edu.tw}
\altaffiltext{7}{National Radio Astronomy Observatory, Charlottesville, VA 22903,
USA; apeck@alma.cl}
\altaffiltext{8}{Department of Astronomy, University of Massachusetts, Amherst, MA 01003;myun@astro.umass.edu}

\begin{abstract}
We present high-resolution ($\sim$ 2.5\arcsec) observations of \cosix\ towards the luminous infrared galaxy VV 114 using the Submillimeter Array. We detect \cosix\ emission from the eastern nucleus of VV 114 but do not detect the western nucleus or the central region. We combine the new \cosix\ observations with previously published or archival low-$J$ CO observations, that include \tcoone\ Atacama Large Millimeter/submillimeter Array cycle 0 observations, to analyze the beam-averaged physical conditions of the molecular gas in the eastern nucleus. We use the radiative transfer code RADEX and a Bayesian likelihood code to constrain the temperature (\tkin), density (\nhtwo) and column density (\nco) of the molecular gas. We find that the most probable scenario for the eastern nucleus is a cold (\tkin\ = 38 K), moderately dense (\nhtwo\ = 10$^{2.89}$ cm$^{-3}$) molecular gas component.  We find the most probable $^{12}$CO to $^{13}$CO abundance ratio (\xco) is 229, roughly three times higher than the Milky Way value. This high abundance ratio may explain the observed high  $^{12}$CO/ $^{13}$CO line ratio ($>$ 25). The unusual $^{13}$CO $J$=2-1/$J$=1-0 line ratio of 0.6 is produced by a combination of moderate $^{13}$CO optical depths ($\tau = 0.4-1.1$) and extremely subthermal excitation temperatures. We measure the CO-to-H$_{2}$ conversion factor, $\alpha_{CO}$ to be 0.5$^{+0.6}_{-0.3}$ \alphaco, which agrees with the widely used factor for ultra luminous infrared galaxies of Downes $\&$ Solomon (1998; $\alpha_{CO}$ =0.8 \alphaco). 

\end{abstract}


\section{Introduction} 
Current ground-based high-resolution observations of molecular gas traced by CO for nearby galaxies are generally limited to the first three rotational transitions ($J$=1-0, $J$=2-1, and $J$=3-2).  The $Fourier$ $Transform$ $Spectrometer$ (FTS) aboard the $Herschel$ $Space$ $Observatory$ has allowed us to observe higher-$J$ transitions of nearby galaxies but at the cost of angular and spectral resolution with most ultra/luminous infrared galaxies (U/LIRGs) being point-like within the FTS beam. The higher-$J$ transitions allow us to trace the hot ($>$ 100 K) molecular gas that is heated by massive young stars, X-ray dominated regions (XDRs; Maloney et al. 1996)\nocite{1996ApJ...466..561M}, shocks \citep{2013ApJ...762L..16M}, low-energy cosmic ray protons (Goldsmith $\&$ Langer 1978) \nocite{1978ApJ...222..881G}or mechanical heating (Rangwala et al. 2011) \nocite{2011ApJ...743...94R}. The Submillimeter Array (SMA; Ho et al. 2004)  \nocite{2004ApJ...616L...1H} has a receiver which provides access to the 690 GHz atmospheric window; however, observations in this window are extremely difficult due to the lack of calibrators and the poor atmospheric transmission. Thus, high angular resolution observations of higher $J$-level CO transitions have been extremely challenging to date.  Observations with the SMA, however, provide a preview to the type of observations that will be possible with the larger collecting area of the Atacama Large Millimeter/submillimeter Array (ALMA).

Over the past two decades several starbursts have been observed in \cosix\ using single dish telescopes (e.g Papadopoulos et al. 2012 and references within).\nocite{2012ApJ...751...10P} Some of the first detections of \cosix\ were towards the starbursts NGC 253, IC 342 and M82 (Harris et al. 1991) \nocite{1991ApJ...382L..75H}where it was shown that the molecular gas in starbursts is warmer than in normal disk galaxies. The only ULIRG to be observed in high-resolution \cosix\ emission, to date, has been Arp 220 (Matsushita et al. 2009) \nocite{2009ApJ...693...56M}using the SMA. Both nuclei of Arp 220 were detected in \cosix\ and 435 $\mu$m continuum emission. Matsushita et al. (2009) find that the nuclei of Arp 220 have very similar $^{12}$CO spectral line energy distributions (SLEDs) to M82 implying similar molecular gas excitation conditions and a density of $\geq$ 10$^{3.3}$ cm$^{-3}$ or a temperature of $\geq$ 30 K. However, using the CO SLED from CO $J$ = 1-0 to CO $J$ = 13-12, Rangwala et al. (2011) find an H$_{2}$ component that is significantly warmer (1350 K) than the 500 K component seen in M82 by Kamenetzky et al. (2012).

VV 114 (Arp 236, IC 1623, IRAS 01053-1746) is a nearby LIRG at a distance of 86 Mpc ($H_{o}$ = 70.5 km s$^{-1}$ Mpc$^{-1}$, $\Omega_{M}$ =  0.27, $\Omega_{\Lambda}$ =  0.73; 1\arcsec = 417 pc) with a far-infrared luminosity of \lfir = 3.16 $\times$ 10$^{11}$ \lsol\ (Sanders et al. 2003)\nocite{2003AJ....126.1607S}. The two merging galaxies have a projected separation of 15\arcsec\ (6.3 kpc; Frayer et al. 1999)\nocite{1999AJ....118..139F}. U et al. (2012) \nocite{2012ApJS..203....9U}have derived a dust temperature of 31.45 $\pm$ 0.15 K and a dust mass of  2.6 $\times$ 10$^{7}$ \msol\ by fitting a spectral energy distribution (SED) to the entire galaxy. \cite{1994ApJ...430L.109Y} observed VV 114 in \coone\ and found large amounts of molecular gas within a bar-like morphology. \cite{2002A&A...391..417L} found evidence of an active galactic nucleus (AGN) associated with the eastern galaxy (VV 114E), which is the brighter of the two nuclei in the near-infrared \citep{2000AJ....119..991S} but highly obscured in the optical \citep{1994AJ....107..920K}. Iono et al. (2013)\nocite{2013PASJ...65L...7I} have shown that VV 114E has a high HCN/HCO$^{+}$ $J$=4-3 line ratio which further supports an AGN within VV 114E. 

VV 114 is one of fourteen U/LIRGs in the SMA sample of Wilson et al. (2008; hereafter Paper I)\nocite{2008ApJS..178..189W}. Compared to submillimeter galaxies (SMGs), local U/LIRGs are more compact (0.3 - 3.1 kpc) than SMGs (3 - 16 kpc; Iono et al. 2009; Paper II). \nocite{2009ApJ...695.1537I}In an earlier paper, \nocite{2012ApJ...753...46S} (Sliwa et al. 2012, Paper III), we used the radiative transfer code, RADEX \citep{2007A&A...468..627V} to constrain the physical conditions of the molecular gas in the three regions of Arp 299. In this paper, we present high-resolution observations of \cosix\ of VV 114 using the SMA. In Section 2, we present the observations and the data reduction process of \cosix, and the recovery of the short spacings of the SMA \cotwo\ and \cothree\ maps published in Wilson et al. (2008).  In Section 3, we present line ratio maps of VV 114. In Section 4, we discuss the \cosix\ emission of VV 114. In Section 5, we present our radiative transfer modeling results. In Section 6, we discuss the implications of the low $^{13}$CO abundance and use our modeling results to constrain the CO-to-H$_{2}$ conversion factor in VV 114.
\section{Observations and Data Reduction} 
\subsection{\cosix\ Submillimeter Array Data} 
The \cosix\ emission in VV 114 was observed with the SMA on 2008 August 26 using a two-point mosaic. Although eight antennas were available, we used only the six antennas that were placed in the innermost ring of the subcompact configuration to reduce resolution effects (i.e. avoid resolving out flux). This resulted in baselines ranging between 10~m and 25~m which corresponds to an angular resolution of 3.1$\arcsec$$\times$2.2$\arcsec$. The upper sideband (USB) of the 690~GHz receiver was tuned to the redshifted \cosix\ line at 677.870~GHz leaving the lower sideband (LSB), separated by 10~GHz, for continuum measurements. Each sideband had a width of 2~GHz with a spectral resolution of 0.8125~MHz.  Weather conditions were excellent throughout the track with an opacity at 225~GHz around 0.02 - 0.06 corresponding to a precipitable water vapour of $\sim$0.5 - 1.0~mm. This corresponds to system temperatures below 600~K. 

Data reduction was done with the IDL MIR package. The bandpass was calibrated on Uranus, Callisto and 3C454.3.  The time variable gain calibration was performed on Uranus which was within 20~degrees of VV 114 at the time of the observations. Uranus had a diameter of 3.6" on the observation date and thus was slightly resolved. We used a standard planet visibility model (a uniform disk) in analyzing the Uranus $uv$-data. The absolute flux scale was set using Uranus and checked using Callisto suggesting an accuracy level of the flux calibration of 20\%. The calibrated $uv$-data were converted to FITS format and then imported into CASA for further processing and imaging. The $uv$-data were flagged to remove outlier data points with high amplitude values and the  continuum was subtracted using line-free channels. We created a clean data cube with velocity resolution of 20 \kms\ using a robust weighting scheme. The dataset was cleaned down to two times the rms noise level of 270 mJy beam$^{-1}$ in each 20 km s $^{-1}$ velocity channel. An integrated intensity map was created using channels with emission greater than the 2$\sigma$ cutoff. The integrated intensity map has been corrected for the primary beam (Figure 1). The peak spectrum of VV 114E is shown in Figure 2. 

\subsection{Additional CO Data}
VV 114 was observed at the SMA in \cothree\ on 2006 November 12 and \cotwo\ on 2006 November 24. The observations and calibration details are given in Paper I. The \tcotwo\ line was observed in the LSB simultaneously with \cotwo. The calibrated $uv$-data from Paper I, in FITS format, were imported into CASA for further processing and imaging. The datasets were processed similarly to the \cosix\ data set (see above and Figures 15 and 16 of Paper I). 

We have obtained a natural weighted clean cube of \coone\ observed with the Owens Valley Radio Observatory (OVRO) first published in Yun et al. (1994). Details of the observations and reduction are given in Yun et al. (1994). We created an integrated intensity map with a 2$\sigma$ cutoff using velocity channels with emission from VV 114. 
VV 114 was observed with ALMA during cycle 0 in \tcoone\ (Saito et al. in prep.) using the compact configuration. We obtained calibrated $uv$-data from the ALMA archive. The $uv$-data were continuum subtracted using line-free channels and a 40 km s$^{-1}$ cleaned datacube was created. We created an integrated intensity map with a 2$\sigma$ cutoff using velocity channels with emission from VV 114 (see Figure 1). 

All integrated intensity maps were corrected for the primary beam. Table 1 presents the observational information for VV 114.

The James Clerk Maxwell Telescope (JCMT) was used to map VV 114 in \cothree\ using the  RxB3 receiver on 2005 August 17 (Program: M05AC05; PI: C.D. Wilson) and \cotwo\ using the RxA3 receiver on 2012 November 26 (Program: M12BC14; PI: K. Sliwa). The \cothree\ map covered 35$\arcsec$ on each side with a 14$\arcsec$ beam and the \cotwo\ map covered 49$\arcsec$ on side with a 21$\arcsec$ beam which included all the emission from VV 114. For the \cothree\ reduction refer to Paper I. The raw \cotwo\ data were made into a cube spanning -600 km s$^{-1}$ to 600 km s$^{-1}$ using the \verb=Starlink= software (Currie et al. 2008)\nocite{2008ASPC..394..650C}. Using line-free channels, the cube was continuum corrected using a first-order baseline. The velocity resolution of the cube was binned to 20 km s$^{-1}$. We assume a main beam efficiency ($\eta_{mb}$) of 0.6 for \cothree\ and 0.69 for \cotwo\ in order to convert the intrinsic units of antenna temperature ($T_{A}^{*}$) to main beam temperature ($T_{mb}$ = $T_{A}^{*}$/$\eta_{mb}$) and then to Jy km s$^{-1}$ which gives us a scaling factor of 26.9 Jy K ($T_{mb}$)$^{-1}$ and 22.9 Jy K ($T_{mb}$)$^{-1}$ for \cothree\ and \cotwo\, respectively. Similarly to Sliwa et al. (2012) with Arp 299, we assume that emission from VV 114 fills the beam, unlike Paper I where VV 114 was treated like a point source. 

When we compare the SMA-only map to the JCMT map, the \cothree\ map is missing about 60$\%$ and the \cotwo\ map about 46$\%$ of the total flux (see Table 1). Note that Paper I estimated only 48$\%$ for \cothree\ because they assumed VV 114 to be point-like in the K to Jy conversion of the JCMT map.

We recover the short spacings of the \cotwo\ and \cothree\ SMA datasets using the JCMT datasets. The recovery of short spacings using single dish data is still a process under development and it will become a common practice in the ALMA era. The two datasets were combined in the image plane using the ``feathering" technique (Stanimirovic 2002)\nocite{2002ASPC..278..375S}. We use the the method of Paper III assuming the flux scales of both datasets to be the same. Figure 1 presents the integrated intensity maps of \cotwo\ and \cothree\ with short spacings recovered. For comparison, see Figures 15 and 16 of Paper I for the SMA-only integrated intensity maps. Both short spacings recovered maps show significant increases in flux (see Figure 1 and Table 1). Negative flux (dashed contours in Figure 1), usually a sign of the missing flux problem, is still seen in the \cothree\ map; however,  negative flux can also be caused by the $uv$-coverage of the map (Stanimirovic 1999)\nocite{1999PhDT........21S}. Since the negative flux caused by the $uv$-coverage cannot be quantified, adjusting the flux scales of the datasets may introduce artificial emission into our maps. We, therefore, attribute the negative flux seen to the poor $uv$-coverage of the SMA maps that transfer into the short spacings recovered maps.

\section{Line Ratio Maps} 
We have degraded the resolution of all the maps to the resolution of the OVRO \coone\ map, 6.5\arcsec $\times$ 3.7\arcsec. This ensures that we are probing the molecular gas of VV 114 on similar physical scales ($\sim$ 2 kpc). We applied a Gaussian taper to the observations to degrade the resolution. We created new integrated intensity maps and corrected for the primary beam. The intrinsic units of the CO integrated maps were converted from Jy beam$^{-1}$ km s$^{-1}$ to K km s$^{-1}$. Line ratio maps of $^{12}$CO ($J$=6-5)/($J$=3-2), $^{12}$CO ($J$=3-2)/($J$=2-1), $^{12}$CO ($J$=2-1)/($J$=1-0), $^{13}$CO ($J$=2-1)/($J$=1-0), ($^{12}$CO/$^{13}$CO) $J$=1-0 and ($^{12}$CO/$^{13}$CO) $J$=2-1 were created by dividing the appropriate maps with a 2$\sigma$ cutoff for each map (Figures 3 and 4). Emission near the edges of the maps was masked out. Note that the line ratio values at the edges of VV 114 are noisier because the uncertainty approaches $\pm$50$\%$. Table 2 presents the line ratios at the peak position of \cothree\ emission in each of the three regions. 

In Paper III, it was found that at the peak position for each nucleus and the overlap region of Arp 299, the missing flux was insignificant ($<<$1$\%$) in both the \cotwo\ and \cothree\ maps. This is not the case for VV 114 and indicates that the short spacings are important in this galaxy. However, when we compare the intensity of the short spacings corrected maps to the SMA-only maps near the peak position of VV 114E, we see that only 10-15$\%$ of flux is missing in the \cothree\ and \cotwo\ maps, well within our calibration uncertainty. Thus, the line ratios at VV 114E are roughly similar whether or not the short spacing data are included (see Figure 3 and Table 2), and we assume the same for the other ratio maps around the peak position of VV 114E. 

\section{\cosix\ Emission}
The \cosix\ line is only detected in VV 114E (Figure 1 and 2). The other two sources $>$3$\sigma$ in the map do not have any counterpart in the \cothree\ map or spectrum, and are thus likely to be artifacts. The total flux of \cosix\ in the SMA map is 150 $\pm$ 40 Jy km s$^{-1}$. Since there are no \cosix\ single dish maps of VV 114, we use an archival $Herschel$ FTS spectrum to estimate the missing flux. The total flux of the \cosix\ line is 1390 $\pm$ 100 Jy km s$^{-1}$ from the FTS spectrum (M.R.P. Schirm 2013, private communication) and when compared to the SMA map, we are missing 89 $\pm$ 10 $\%$ of the flux. The missing flux indicates that short spacings are increasingly important at higher observing frequencies for VV 114.

If we assume the same $^{12}$CO($J$=6-5)/($J$=3-2) line ratio for the central (VV 114C) and western (VV 114W) regions as for VV 114E, we can estimate the expected peak intensity of \cosix\ of the other two regions. This estimate yields 130 and 70 Jy beam$^{-1}$ km s$^{-1}$ for VV 114C and VV 114W regions respectively and suggests that we should have detected the other two regions at $\geq$4$\sigma$ level. This situation could be explained by two scenarios: the emitting \cosix\ sources in these two regions are on larger scales that are filtered out with the SMA or the $^{12}$CO($J$=6-5)/($J$=3-2) line ratio is quite different from VV 114E indicating different physical conditions. If we assume the former to be true, the $Herschel$ FTS observations cannot be used to recover the missing flux in the SMA map; a single dish \cosix\ map is required to recover the flux. If we assume the latter to be true, we can use the 3$\sigma$ level of the \cosix\ map and the peak intensity in the \cothree\ SMA map to put an upper limit to the $^{12}$CO($J$=6-5)/($J$=3-2) line ratio (see Table 2). 

We compare the \cosix\ maps of VV 114 and Arp 220 (Matsushita et al. 2009). The SMA \cosix\ observations of Arp 220 kinematically resolved the two nuclei. When comparing the SMA \cosix\ flux of Arp 220 with the $Herschel$ $FTS$ flux (Rangwala et al. 2011), the SMA map is missing 69 $\pm$ 7 $\%$. More \cosix\ emission is recovered in Arp 220 with the SMA which suggests that the \cosix\ emission is more compact than in VV 114.  The $^{12}$CO ($J$=6-5)/($J$=3-2) line ratios of the two nuclei of Arp 220 are nearly five times higher than for VV 114E indicating that the molecular gas is more excited in Arp 220 than in VV 114. Rangwala et al. (2011) have found evidence of a very hot (\tkin\ = 1350 K)  molecular gas component which may explain the higher observed $^{12}$CO ($J$=6-5)/($J$=3-2) line ratio.  The $^{12}$CO ($J$=6-5)/($J$=3-2) line ratio for M82 ($\sim$ 0.35; Kamenetzky et al. 2012) is also higher than that of VV 114E. The hot molecular gas component of M82 (\tkin\ = 500 K; Kamenetzky et al. 2012) is cooler than of Arp 220. If the differences in the  $^{12}$CO ($J$=6-5)/($J$=3-2) line ratios are due to temperature differences, then the hot molecular gas component of VV 114E would be less than 500 K.

\section{Molecular Gas Physical Conditions}
\subsection{Radiative Transfer Model} 
In order to constrain the beam-averaged physical conditions of VV 114, we use the radiative transfer code RADEX (van der Tak et al. 2007). RADEX calculates line fluxes and optical depth for specific temperature (\tkin), density (\nhtwo) and column density (\nmol) of the molecular species of interest. 

We use RADEX to generate $^{12}$CO line fluxes for a 3D grid in \tkin\ (10$^{0.7}$ - 10$^{3.8}$ K), \nhtwo\ (10$^{1.0}$ - 10$^{7.0}$ cm$^{-3}$), and \nco\ (10$^{12}$ - 10$^{18}$ cm$^{-2}$) for $dv$ = 1 km s$^{-1}$. We use the line widths measured using the \cothree\ map (see Table 3) to correct the column density for the particular region of interest in VV 114. The observed Galactic $^{13}$CO abundance relative to $^{12}$CO is \xco\ = 30 -- 70 (Langer $\&$ Penzias 1990)\nocite{1990ApJ...357..477L}. We generate another grid for $^{13}$CO using the same parameter space as for $^{12}$CO that covers the observed Galactic abundance value range (\xco\ =10 -- 100; Grid 1). We also generate a second $^{13}$CO grid expanding the relative abundance to $^{12}$CO (\xco\ = 10 -- 10$^{5}$; Grid 2). 


\subsection{Likelihood Analysis}
In addition to RADEX, we use a Bayesian likelihood code (Ward et al. 2003, Panuzzo et al. 2010, Kamenetzky et al. 2011)\nocite{2010A&A...518L..37P}\nocite{2003ApJ...587..171W}\nocite{2011ApJ...731...83K}. The Bayesian likelihood code determines the most probable solutions for the physical conditions of the molecular gas by comparing the RADEX  calculated and the observed $^{12}$CO and $^{13}$CO line fluxes. The likelihood code generates probability distributions for four parameters,  \tkin ,  \nhtwo ,  \nco\ and area filling factor (\ff). We mention some of the basic aspects of the code here but for more details on the likelihood code see Kamenetzky et al. (2012)\nocite{2012ApJ...753...70K} and Rangwala et al. (2011). 

The likelihood code generates two calculated values for each parameter: the 1DMax and the 4DMax. The 1DMax refers to the maximum likelihood of the given parameter based on the one-dimensional likelihood distribution for the parameter. The 4DMax is the maximum likelihood of a single grid point based on the four-dimensional likelihood distribution of the four parameters mentioned above. 

We implement three priors in order to constrain the solutions to those of realistic physical conditions (Ward et al. 2003; Rangwala et al. 2011). The first prior constrains the column density to ensure that the total mass within the column does not exceed that of the dynamical mass of the system (see Rangwala et al. 2011 for more details). The dynamical mass of each region was measured using
\begin{equation}
M_{\rm{dyn}} = 99\Delta V^{2} D\rm{(pc)}
\end{equation}
where $\Delta V$ is the line width of the system in km s$^{-1}$ and $D$(pc) is the diameter of the system in parsecs (Wilson et al. 1997; see Table 3)\nocite{1997ApJ...483..210W}. The \cothree\ map has the best resolution and signal-to-noise ratio (SNR); therefore, we use the \cothree\ map to fit each region of VV 114 to a two-dimensional Gaussian to measure the diameter (see Table 3), assuming a spherical geometry. We also use the \cothree\ spectrum of each region to fit $\Delta V$ using a Gaussian profile (see Table 3). 

The second prior constrains the column length to be less than the length of the molecular region. We estimate the length of the molecular region using the deconvolved source sizes (see Table 3) and assuming a spherical geometry. This prior will constrain \nhtwo\ at the lower end and \nco\ at the higher end (Rangwala et al. 2011). 

The third prior constrains the optical depth to be 0 $<$ $\tau$ $<$ 100. A negative $\tau$ indicates a maser; however, we do not expect the CO lines to exhibit maser properties and, therefore, we constrain $\tau$ to be greater than zero. The upper limit of $\tau$ = 100 is put in place as recommended by the RADEX documentation. 

\subsection{Modeling Results}
We model the molecular gas at the peak position of \cothree\ ($\alpha_{J2000}$ = 01$^{h}$07$^{m}$47.453$^{s}$,  $\delta_{J2000}$ = -17$^{d}$30$^{m}$25.037$^{s}$) in VV 114E. We do not have detections of \cosix\ and $^{13}$CO in VV 114W; therefore, we cannot model the molecular gas of VV 114W using the likelihood code (i.e. four fitted parameters using three observed line fluxes).  

We present two possible solutions for VV 114E, one for each of the $^{13}$CO grids. The most probable solution for Grid 1 gives a hot (\tkin\ = 390 K), diffuse (\nhtwo\ = 10$^{2.29}$ cm$^{-3}$) molecular gas component. The most probable \xco\ value hit the upper limit of our grid (\xco\ = 100). To see if hitting the upper limit of \xco\ affects our results, we use Grid 2 with the expanded range of values. Grid 2 shows a maximum probable solution of a cold (\tkin\ = 38 K), moderately dense (\nhtwo\ =  10$^{2.89}$ cm$^{-3}$) molecular gas component (see Table 4). The most probable \xco\ value increased from 100 to 229. The difference in the solutions arises from the difference in the most probable \xco\ value showing the importance of the abundance ratio in constraining the physical conditions of the molecular gas. Figure 5 shows the $^{12}$CO and $^{13}$CO SLEDs with the most probable solution fit to both species. Figure 6 shows the one-dimensional probability distributions for \tkin\, \nhtwo\ and \nco\ for both grid runs to show the difference in the solutions. Figure 7 shows the two-dimensional distribution for \tkin\ against \nhtwo\ for Grid 2. The optical depth ($\tau$) for the $^{12}$CO lines is $>>$1 where $\tau$ $\sim$ 14 for \coone\ and peaks at the $^{12}$CO $J$=4-3 line with $\tau$ $\sim$ 65 indicating that the $^{12}$CO lines are optically thick. The $\tau$ for $^{13}$CO peaks at the \tcotwo\ line with $\tau$ $\sim$ 1.1 and the \tcoone\ $\tau$ $\sim$ 0.4.

\section{Discussion}
\subsection{Line Ratios}
Aalto et al. (1995)\nocite{1995A&A...300..369A} show that $^{13}$CO ($J$=2-1)/($J$=1-0) $>$ $^{12}$CO ($J$=2-1)/($J$=1-0) $>$ 1 for the majority of their sample of infrared-bright galaxies. VV 114E shows the opposite with $^{12}$CO ($J$=2-1)/($J$=1-0) $>$ $^{13}$CO ($J$=2-1)/($J$=1-0) at the \cothree\ peak position (see Table 2).  From the RADEX models, we find the excitation temperatures ($T_{\rm{ex}}$) for \tcoone\ and \tcotwo\ are $\sim$14 K and $\sim$8 K, respectively. Assuming similar $T_{\rm{ex}}$ as found with the RADEX models with $\tau$ $<<$ 1 and using the equations in Section 4.2 of the RADEX manual\footnotemark\footnotetext{www.strw.leidenuniv.nl/$\sim$moldata/radex$\_$manual.pdf}, we would expect to see $^{13}$CO ($J$=2-1)/($J$=1-0) $\sim$ 1.4. If both \tcoone\ and \tcotwo\ lines had the same $T_{\rm{ex}}$ and the moderate optical depths as found with the likelihood analysis, we would expect a line ratio $>$ 1.5. This analysis suggests that both moderate optical depths and subthermal excitation are required to produce $^{13}$CO ($J$=2-1)/($J$=1-0) $<$ 1 in VV 114E.

\subsection{Radiative Transfer Modeling Results}

The first three $^{12}$CO transition line intensities are slightly over estimated by the best fit model (see Figure 5; for \cosix\ see below) but the best fit intensities are well within our uncertainties. In order to significantly excite \cothree, the molecular gas must be \tkin\ $>$ 30 K. Once \nhtwo, \tkin\ and \nco\ are large enough to thermalize $^{12}$CO up to $J$=3-2, the $^{13}$CO lines provide further constraints on the best fit.  The $^{13}$CO ($J$=2-1)/($J$=1-0) line ratio will depend on \nthreeco\ and $T_{\rm{ex}}$ (see Section 6.1). A large \nthreeco\ is needed to obtain the moderate optical depths required for the low observed line ratio ($<$ 1). The best fit is likely mainly driven by the $^{13}$CO lines which may explain the slight over-estimate of the $^{12}$CO lines up to $J$=3-2.

The most probable solution for VV 114E is a cold (\tkin\ = 38 K), moderately dense (\nhtwo\ =10$^{2.89}$ cm$^{-3}$) molecular gas component. This solution is similar to what is seen in radiative transfer modeling of Arp 220 (Rangwala et al. 2011), Arp 299 (Paper III) and M82 (Kamenetzky et al. 2012) where a cold, moderately dense molecular gas component is seen. The temperature of the cold molecular gas component in M82 is much warmer (\tkin\ = 63 K) and the gas density is slightly higher (\nhtwo\ = 10$^{3.40}$ cm$^{-3}$) than what we see for VV 114E.  The solution for the physical conditions of VV 114E is more similar to that of the cold molecular gas component in Arp 220 (\tkin\ = 50K, \nhtwo\ = 10$^{2.8}$ cm$^{-3}$). The temperature of the cold molecular gas component in Arp 220 is slightly higher than seen in VV 114E but that may be due to the difference in merger stages of the two galaxies (i.e. Arp 220 is at a more advanced merger stage than VV 114). VV 114 is believed to be a late-stage merger on the verge of experiencing more massive bursts of star formation and becoming a ULIRG (Iono et al. 2004)\nocite{2004ApJ...616L..63I} like Arp 220. The future starburst that VV 114 will experience may heat the cold molecular gas to the same temperature of the cold molecular gas component in Arp 220. 

The $FTS$ analyses of Arp 220 and M82 also shows a hot molecular gas component from excess emission above the $^{12}$CO $J$ = 4-3 transition. From the CO SLED of VV 114E, we do not see any excess emission at the \cosix\ transition; indeed, the 4DMax solution overestimates the \cosix\ flux. The overestimation of the \cosix\ flux may be due to more missing flux than we originally assumed; however, we need a high-resolution image with short spacings to confirm this case. To constrain the hot molecular gas of VV 114 is beyond the scope of this paper, where we require other higher-$J$ level CO transitions at high angular resolution which ALMA will eventually be able to observe. 

\subsection{$^{13}$CO to $^{12}$CO Abundance: \xco}
The most probable value of \xco\ in VV 114E is roughly three times larger than the value seen in the local interstellar medium (ISM; \xco\ = 77; Wilson $\&$ Rood 1994)\nocite{1994ARA&A..32..191W}. Casoli et al. (1992) have suggested that the observed higher $^{12}$CO/$^{13}$CO line ratios seen in U/LIRGs than in our Galaxy are likely due to an under-abundance of $^{13}$CO and/or an over-abundance of $^{12}$CO. Casoli et al. (1992)\nocite{1992A&A...264...55C} also suggest several mechanisms to account for the increased \xco\ value in U/LIRGs with two mechanisms as the most likely scenarios. The starburst episodes of U/LIRGs produce numerous massive stars that can enrich the ISM in $^{12}$CO via $^{12}$C nucleosynthesis that can enhance the \xco\ value. Another possible mechanism to enhance \xco\ is via infalling gas with a high \xco\ value from the outer regions during the merger process.  Both mechanisms can explain our results and it may be a combination of the two. 

Mart\'{\i}n  et al. (2010)\nocite{2010A&A...522A..62M} have placed limits on the $^{12}$C/$^{13}$C isotopic ratio in M82 and NGC 253 to be $>$ 138 and $>$ 81, respectively. If we assume that the  $^{12}$C/$^{13}$C isotopic ratio reflects  \xco\ then both these starbursts have larger \xco\ value than seen the local ISM. Henkel et al. (2010)\nocite{2010A&A...516A.111H} have observed $^{13}$CO $J$=3-2 in the Cloverleaf quasar and have compared it to the \cothree. The observed $^{12}$CO/$^{13}$CO line ratio was 40 and exceeds that seen in Arp 220 but is similar to that of NGC 6240 (Greve et al. 2009)\nocite{2009ApJ...692.1432G}. Henkel et al. (2010) surmise that the high line ratio may be caused by a real deficiency in $^{13}$CO where Large Velocity Gradient (LVG) models yield \xco\ values $>>$ 100. It may be quite common to find larger \xco\ values in starbursts than the value seen in the Galaxy. 

\subsection{CO-to-H$_{2}$ Conversion Factor: $\alpha_{CO}$}
Using the 4DMax solution for the beam-averaged column density ($<$\nco$>$) and assuming a $^{12}$CO to H$_{2}$ abundance ratio of \xh\ = 3 x 10$^{-4}$ (Ward et al. 2003)\nocite{2003ApJ...587..171W}, we can measure the CO-to-H$_{2}$ conversion factor ($\alpha_{CO}$). We use the short spacing corrected \cotwo\ map to measure the peak luminosity of VV 114E because the map has recovered the flux on all scales. We convert the \cotwo\ luminosity to a \coone\ luminosity using the line ratio in Table 2. The total mass within the beam is 6.2$^{+7.3}_{-3.7}$ $\times$ 10$^{8}$ \msol\ and the luminosity of \coone\ within the beam is 1.3 $\times$ 10$^{9}$ K km s$^{-1}$ pc$^{2}$. We measure $\alpha_{CO}$ to be 0.5$^{+0.6}_{-0.3}$ \alphaco. This value agree within uncertainties with the value widely used for ULIRGs determined by Downes $\&$ Solomon (1998; $\alpha_{CO}$ = 0.8 \alphaco). To recover the Downes $\&$ Solomon (1998)\nocite{1998ApJ...507..615D} value, the \xh\ value would need to be 1.6 times smaller (\xh\ = 1.8 $\times$ 10$^{-4}$).  This value for $\alpha_{CO}$ for VV 114E is consistent with the recent findings of Papadopoulos et al. (2012) for LIRGs using single dish observations ($\alpha_{CO}$ = 0.6 $\pm$ 0.2  \alphaco). 

We can place another limit to $\alpha_{CO}$ using the dynamical mass (see Table 2) and the total CO luminosity of VV 114E. Because of its better angular resolution, we use the short spacings corrected \cothree\ map to measure the CO luminosity of VV 114E. We assume a line ratio of 0.92 for $^{12}$CO($J$=3-2)/($J$=1-0) to convert the total \cothree\ luminosity to total \coone\ luminosity of VV 114E ($L_{CO(1-0)}$ = 1.6 $\times$ 10$^{9}$ K km s$^{-1}$ pc$^{2}$).  Comparing this luminosity to the dynamical mass, we measure $\alpha_{CO}$ to be 1.5 \alphaco. Since the dynamical mass includes the mass of stars and dark matter as well as gas, and the gas in the center of VV 114E might not be in gravitational equilibrium, the dynamical mass measured $\alpha_{CO}$ is strictly an upper limit. The upper limit is consistent with the radiative transfer model-derived value for $\alpha_{CO}$.

Paper III measured $\alpha_{CO}$ of Arp 299 to be 0.4 $\pm$ 0.3 \alphaco. The value for Arp 299 is consistent with the value of Downes $\&$ Solomon (1998). Arp 299 is believed to be an earlier stage merger than VV 114. Schirm et al. (in prep.) have measured a value for $\alpha_{CO}$ in NGC 4038/39 ($\alpha_{CO}$ = 7 \alphaco) similar to the Galactic value. NGC 4038/39 is a young merger that has not yet progressed into the LIRG stage and shows no strong evidence of an enhanced conversion factor. The $\alpha_{CO}$ factor may evolve as the merger process goes on instead of the currently adopted bimodal factor (i.e. Galactic or ULIRG); however, a larger sample of measured $\alpha_{CO}$ for U/LIRGs at different stages is needed to deduce whether an $\alpha_{CO}$ factor as a function of merger stage is plausible.

\section{Conclusions}
In this paper, we have presented high-resolution SMA observations of \cosix\ of the nearby LIRG VV 114. We combine the new \cosix\ map with the \cotwo, \cothree\ and \tcotwo\ SMA maps of Wilson et al. (2008), \coone\ OVRO map of Yun et al. (1994) and \tcoone\ ALMA cycle 0 map of Saito et al. (in prep.) to constrain the beam-averaged physical conditions of the molecular gas. 
\begin{enumerate}
\item We detect \cosix\ emission in VV 114E. If we assume a similar $^{12}$CO ($J$=6-5)/($J$=3-2) line ratio for both VV 114C and VV 114W and using the peak position flux of \cothree\, we should have detected VV 114C and VV 114W in \cosix\ at $>$4$\sigma$ level. Since we do not detect VV 114C and VV 114W, we conclude that either the physical conditions of these two regions are different from VV 114E or the \cosix\ emission is extended and filtered out by the SMA. 
\item We combine JCMT \cotwo\ and \cothree\ maps with the SMA maps to recover the short spacings. We find that the missing flux around the position of the peak flux of VV 114E in both maps is $\sim$ 10 -- 15$\%$, well within our calibration uncertainties (20$\%$). The missing flux in VV 114C and VV 114W is more significant, revealing the importance of short spacings for these two regions.  
\item We use the radiative transfer code RADEX and a Bayesian likelihood code to constrain the physical conditions of VV 114E. We find evidence of a cold (\tkin\ = 38 K), moderately dense (\nhtwo\ = 10$^{2.89}$ cm$^{-3}$) molecular gas component. The most probable \xco\ for VV 114 was found to be 229. We show that a lower value of \xco\ would support a hot (\tkin\ $>$ 200 K), diffuse (\nhtwo\ $<$ 10$^{2.5}$ cm$^{-3}$) molecular gas component. 
\item The most probable \xco\ is roughly three times larger than seen for the local ISM and may explain the high $^{12}$CO/$^{13}$CO line ratios ($>$ 25) observed for VV 114E. 
\item We use the most probable $<$\nco$>$ and the peak position luminosity of \coone\ to measure the CO-to-H$_{2}$ conversion factor, $\alpha_{CO}$. We measure $\alpha_{CO}$ to be 0.5$^{+0.6}_{-0.3}$ \alphaco\ for VV 114E, which agrees with the widely used factor of Downes $\&$ Solomon (1998). Using the dynamical mass and the total CO luminosity of VV 114E, we place an upper limit to $\alpha_{CO}$ to be $\leq$ 1.5 \alphaco.
\end{enumerate}

ALMA will eventually make observations of VV 114 in \cosix\ that will include the short spacings from the Atacama Compact Array (ACA). Combining the ALMA \cosix\ observations with the new short spacings recovered SMA \cotwo\ and \cothree\ maps will allow us to constrain the physical conditions of the molecular gas over the full extent of the galaxy using a similar approach to this paper. There is also evidence that a hot molecular gas component exists in starburst galaxies (Rangwala et al. 2011, Kamenetzky et al. 2012); however, to constrain the physical conditions of the hot molecular gas component, higher-$J$ CO transitions are required. ALMA will one day be able to observe VV 114 in $^{12}$CO $J$=4-3 (band 8), $J$=7-6 and $J$=8-7 (band 10) at high angular resolutions that will be used to constrain the physical conditions of the hot molecular gas component.

\acknowledgments{
We thank the anonymous referee for the useful comments that greatly improved our analysis. We also thank the Z-Spec team for allowing us to use their likelihood code. The Submillimeter Array is a joint project between the Smithsonian Astrophysical Observatory and the Academia Sinica Institute of Astronomy and Astrophysics and is funded by the Smithsonian Institution and the Academia Sinica. The James Clerk Maxwell Telescope is operated by the Joint Astronomy Centre on behalf of the Science and Technology Facilities Council of the United Kingdom, the National Research Council of Canada, and (until 31 March 2013) the Netherlands Organisation for Scientific Research. This paper makes use of the following ALMA data: ADS/JAO.ALMA$\#$2011.0.00467.S. ALMA is a partnership of ESO (representing its member states), NSF (USA) and NINS (Japan), together with NRC (Canada) and NSC and ASIAA (Taiwan), in cooperation with the Republic of Chile. The Joint ALMA Observatory is operated by ESO, AUI/NRAO and NAOJ. C.D.W. acknowledges support by the Natural Science and Engineering Research Council of Canada (NSERC). K.S. acknowledges support by the Queen Elizabeth II Graduate Scholarship in Science and Technology (QEIIGSST). MJ acknowledges the support of the Academy of Finland grant No. 250741. This research made use of APLpy, an open-source plotting package for Python hosted at http://aplpy.github.com}


\begin{figure}[h] 
\centering
$\begin{array}{c@{\hspace{0.5in}}c}
\includegraphics[ scale=0.3]{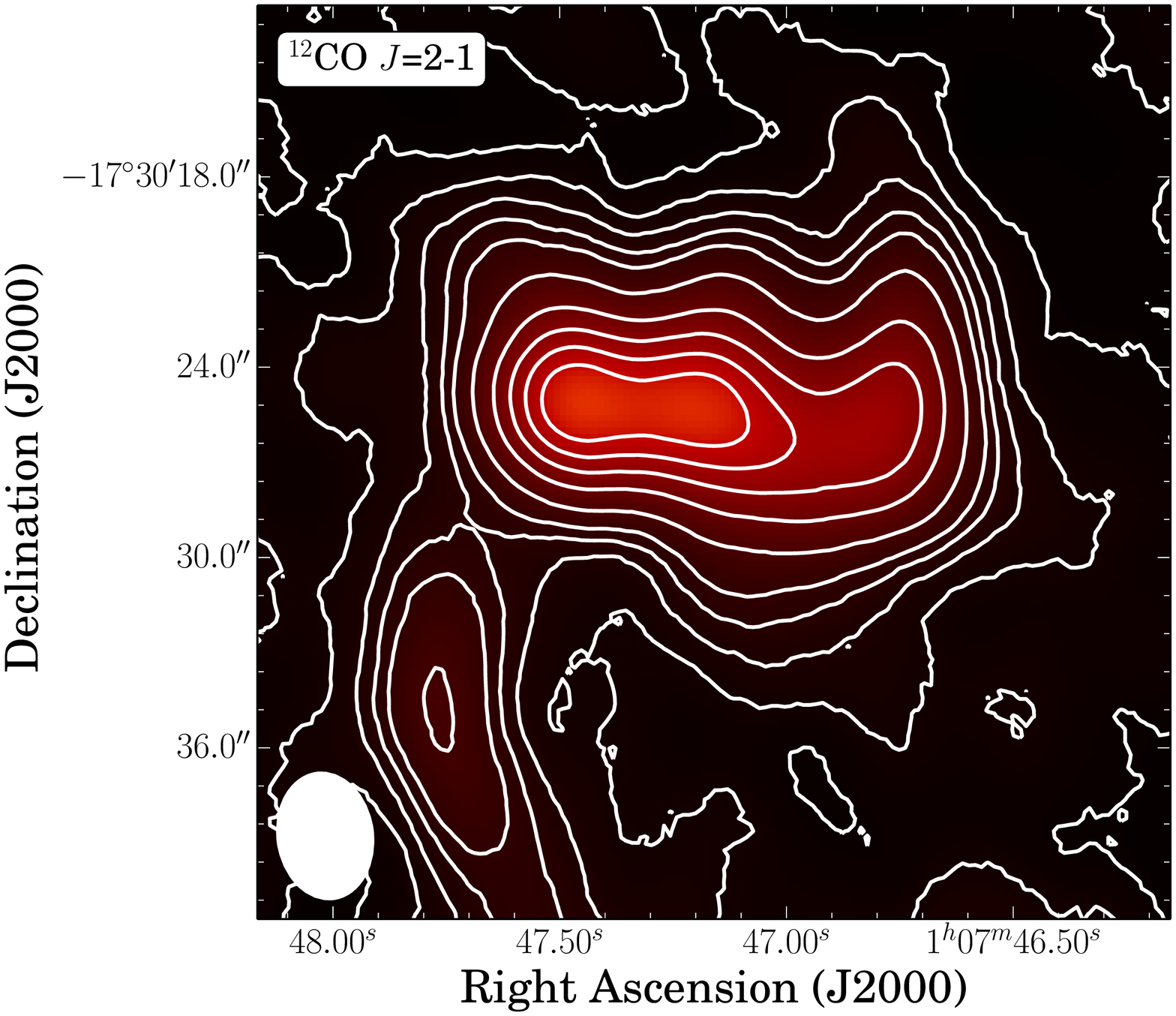}  & \includegraphics[scale=0.3]{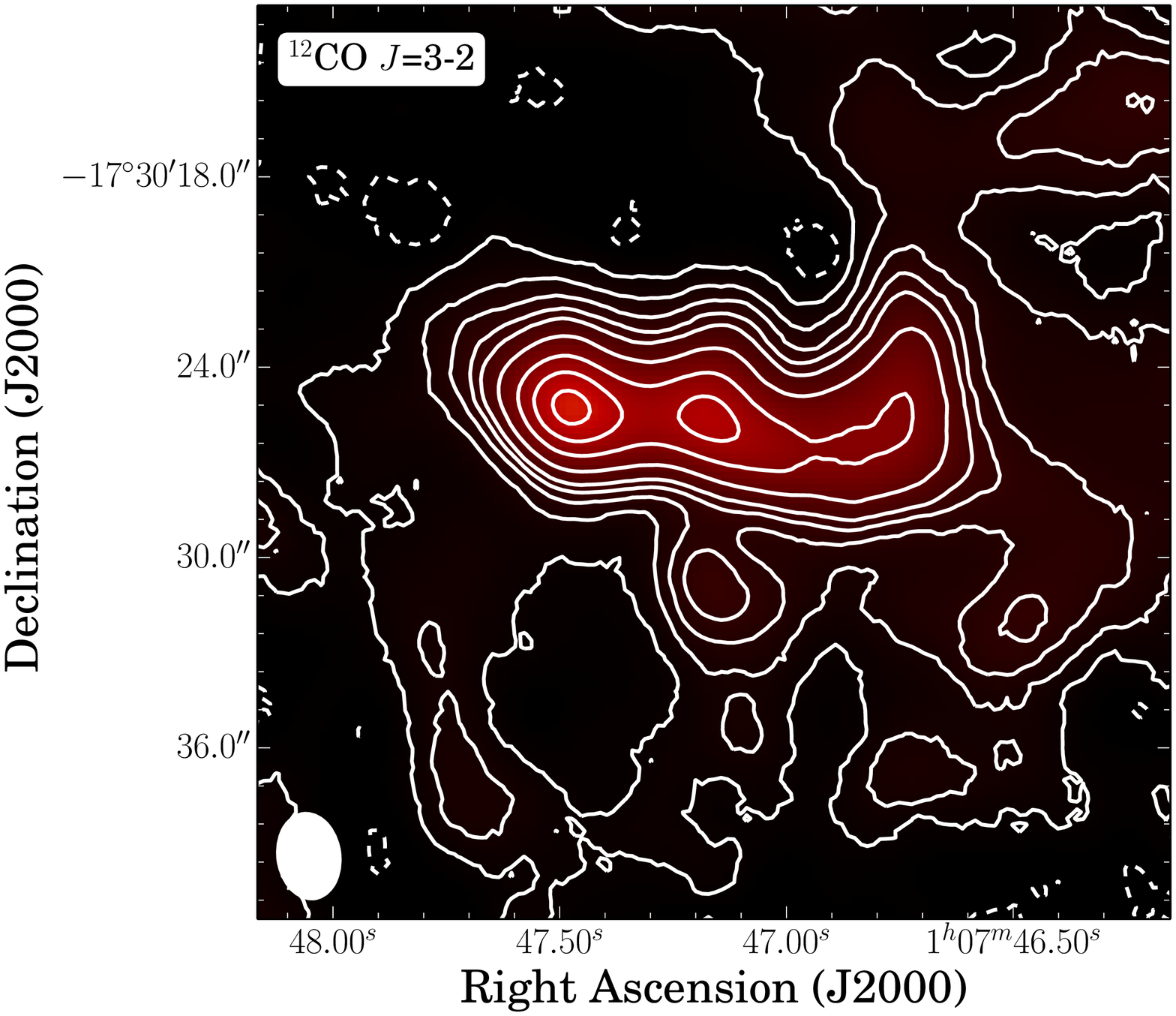} \\
\includegraphics[scale=0.3]{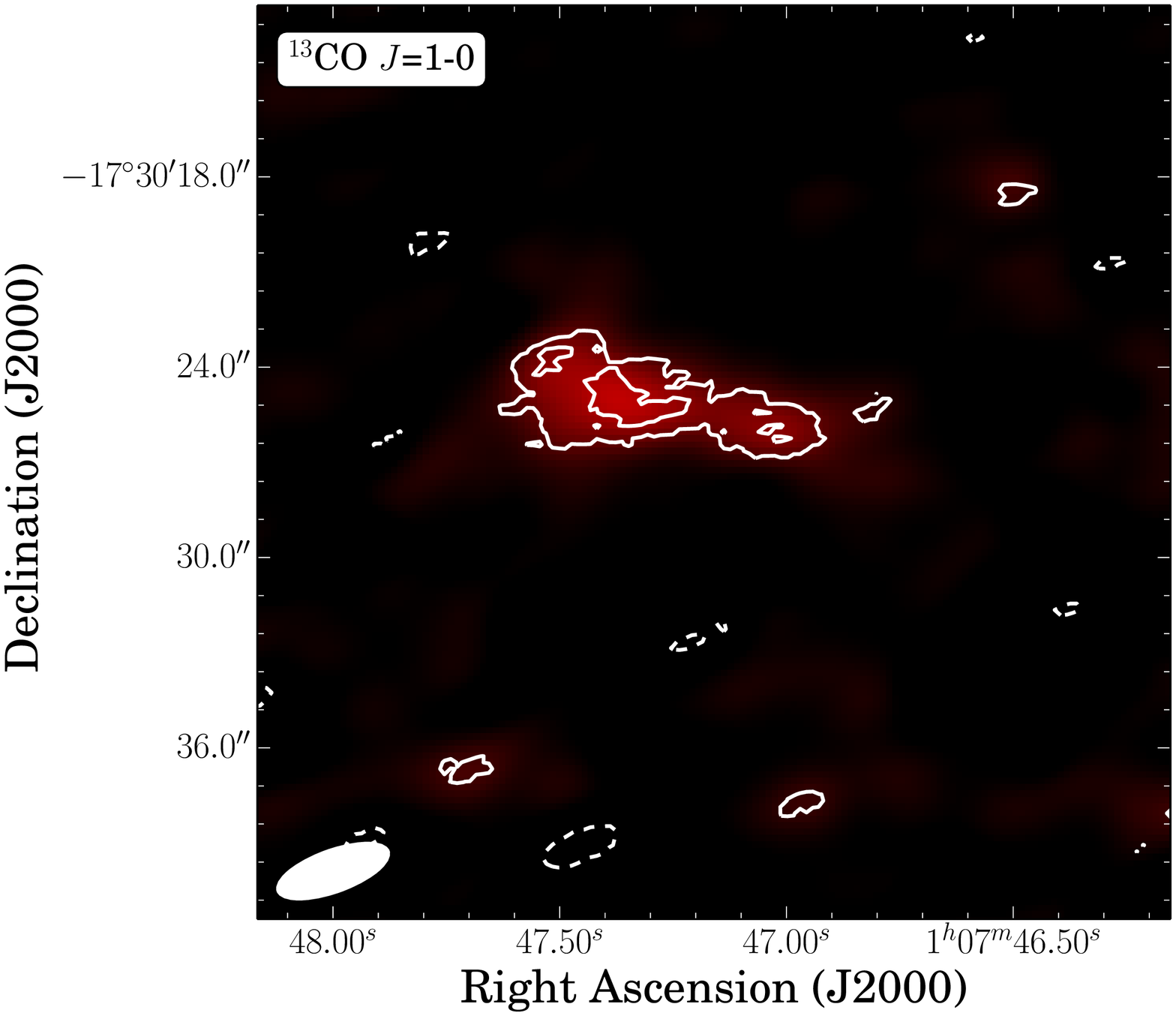} & \includegraphics[scale=0.3]{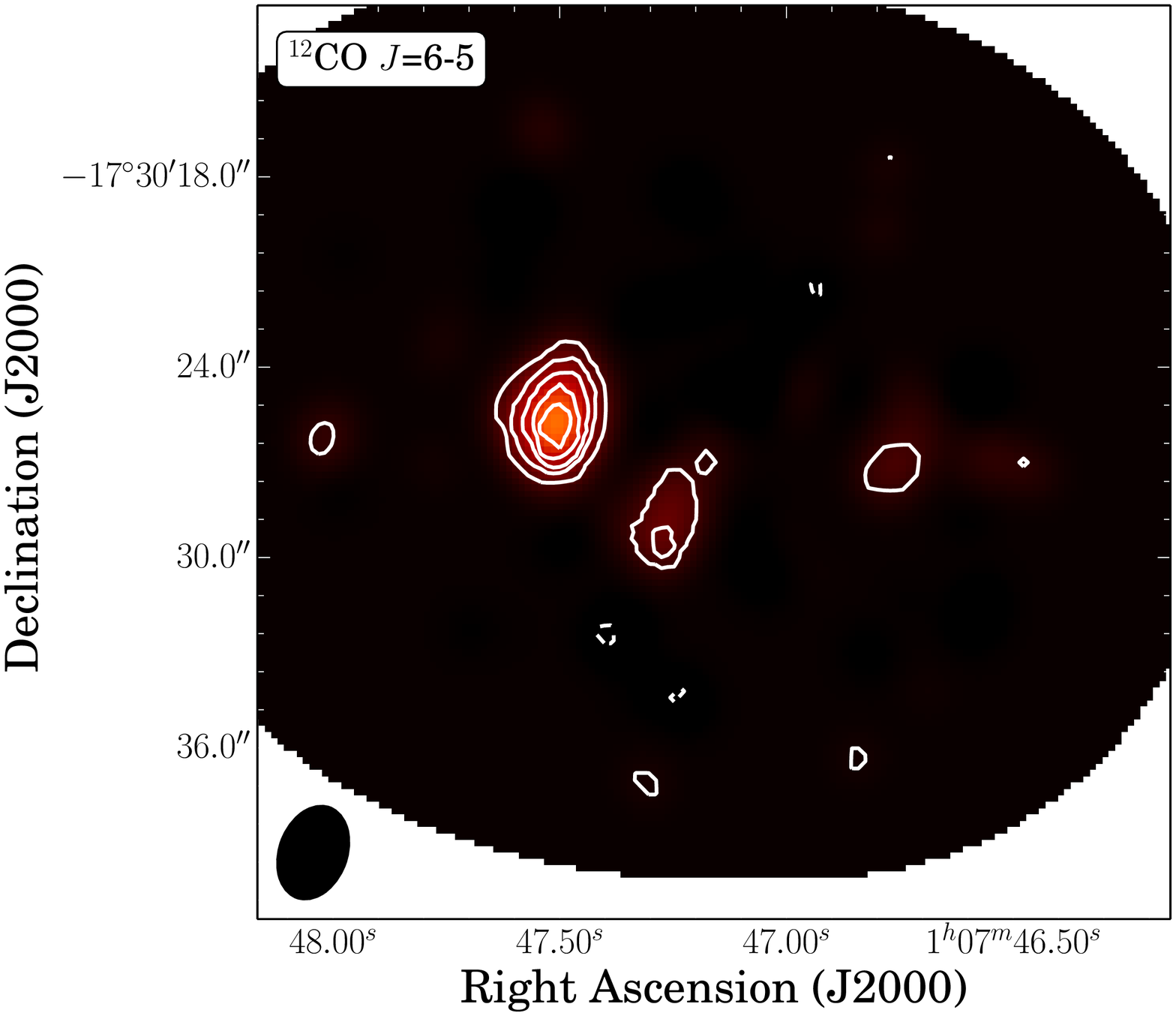} \\
\end{array}$
\caption[]{(\textit{Top Left }) $^{12}$CO J=2-1 map with the recovered short spacings from the JCMT. The contours correspond to -7.5, -4.5, -1.5, 1.5, 4.5, 7.5, 10.5, 13.5, 20, 30, 40, 50, 60 $\times$ 3.5 Jy beam$^{-1}$ km s$^{-1}$ (\textit{Top Right}) $^{12}$CO J=3-2 map with the recovered short spacings from the JCMT. The contours correspond to  -7.5, -4.5, -1.5, 1.5, 4.5, 7.5, 10.5, 13.5, 20, 30, 40, 50 $\times$ 7 Jy beam$^{-1}$ km s$^{-1}$. (\textit{Bottom Left}) \tcoone\ map with contours corresponding to -2, 2, 4, 6 $\times$ the 1$\sigma$ value of 0.25 Jy beam$^{-1}$ km s$^{-1}$. (\textit{Bottom Right}) \cosix\ map with contours corresponding to  --2, 2, 4, 6, 8, 10 $\times$ the 1$\sigma$ value of 16.4 Jy beam$^{-1}$ km s$^{-1}$. All maps have been corrected for the primary beam. Negative flux is denoted by dashed contours. }
\label{SMAmaps}
\end{figure}

\begin{figure}[h] 
\centering
\includegraphics[scale=0.5,angle=-90]{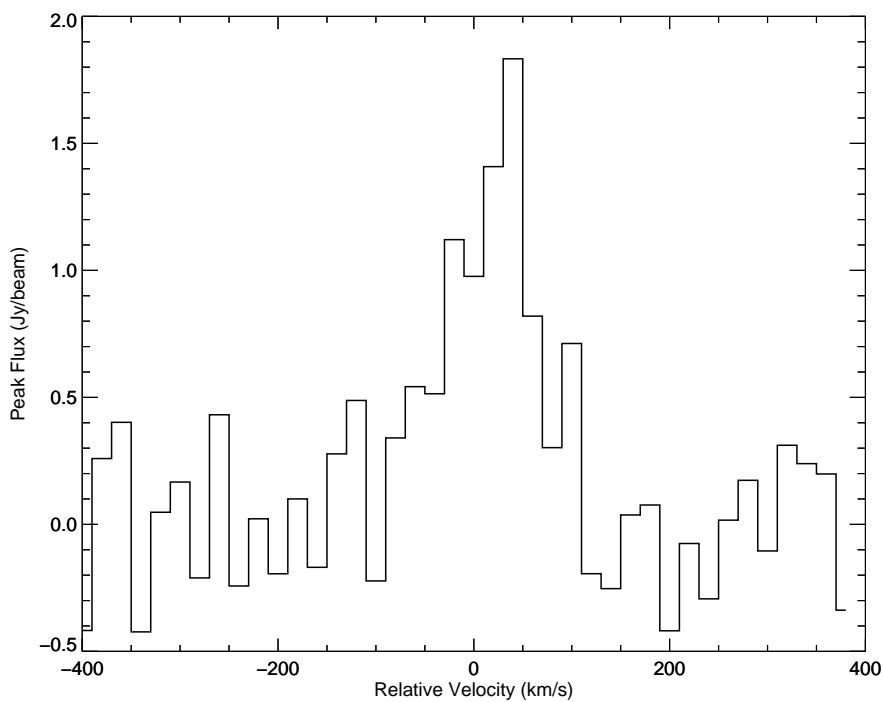} 
\caption[]{Peak spectrum of the \cosix\ emission for VV 114E.}
\label{Inte}
\end{figure}

\begin{figure}[h] 
\centering
$\begin{array}{c@{\hspace{0.5in}}c}
\includegraphics[scale=0.35]{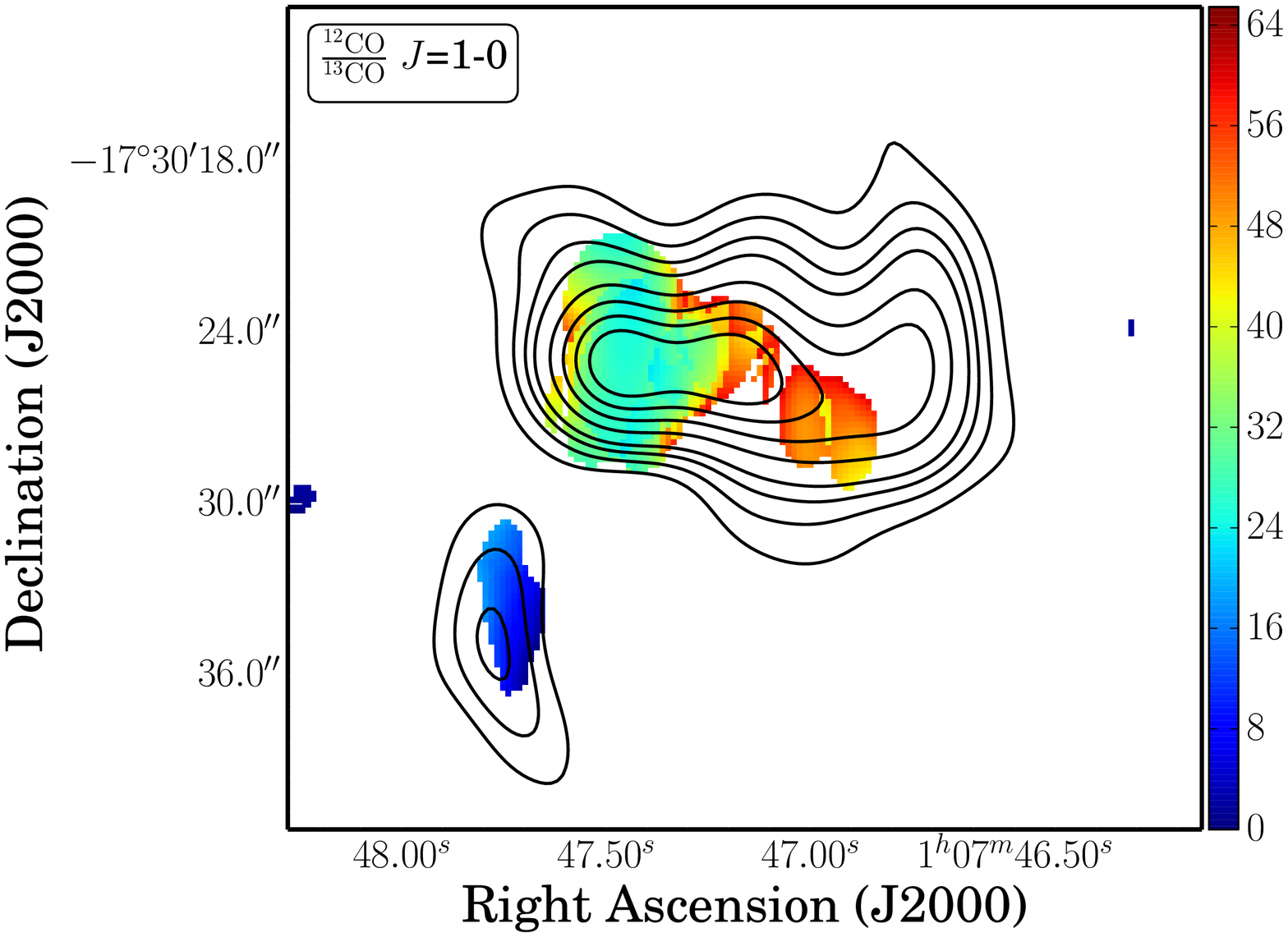} & \includegraphics[scale=0.35]{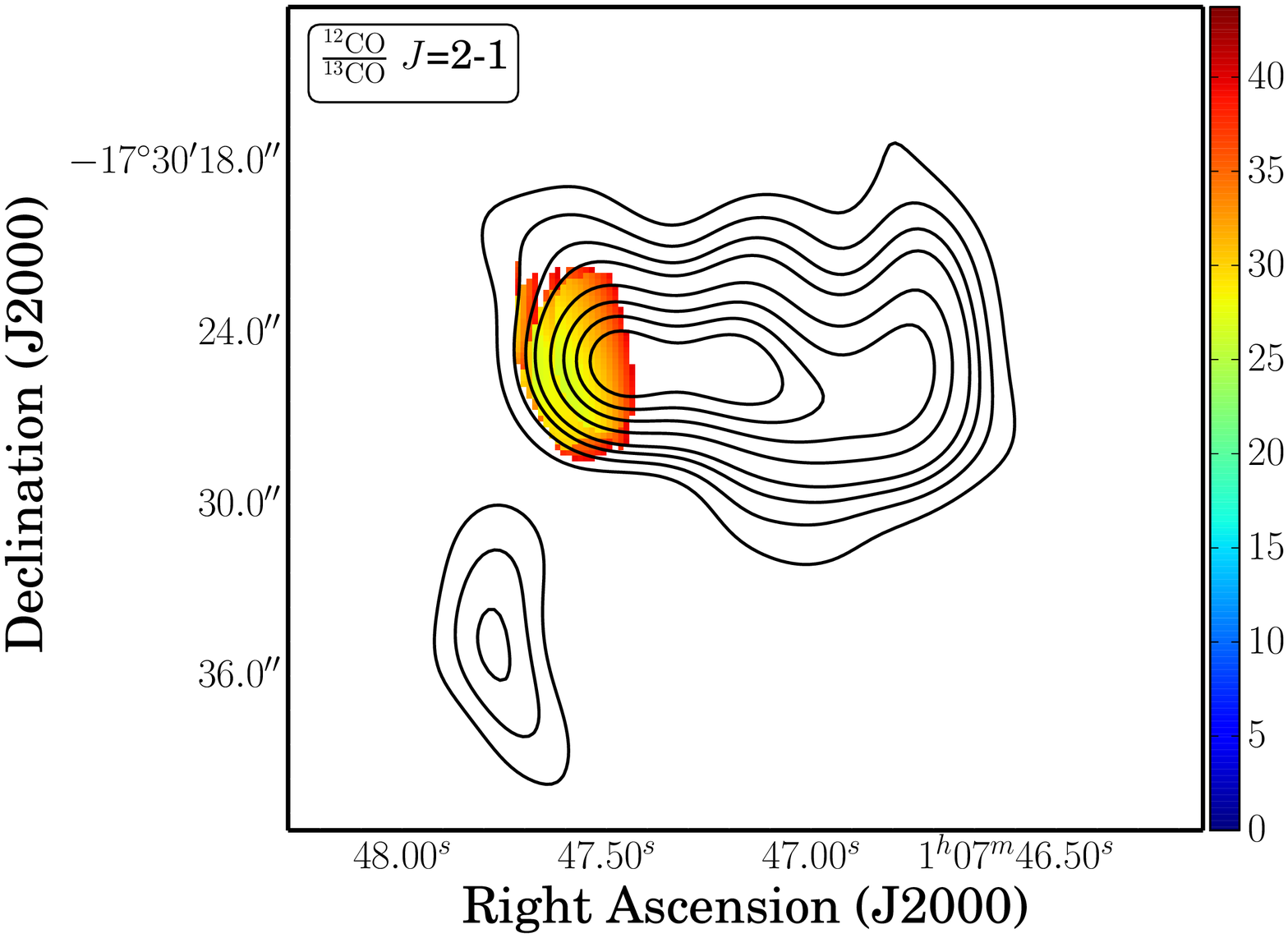} \\
\includegraphics[scale=0.35]{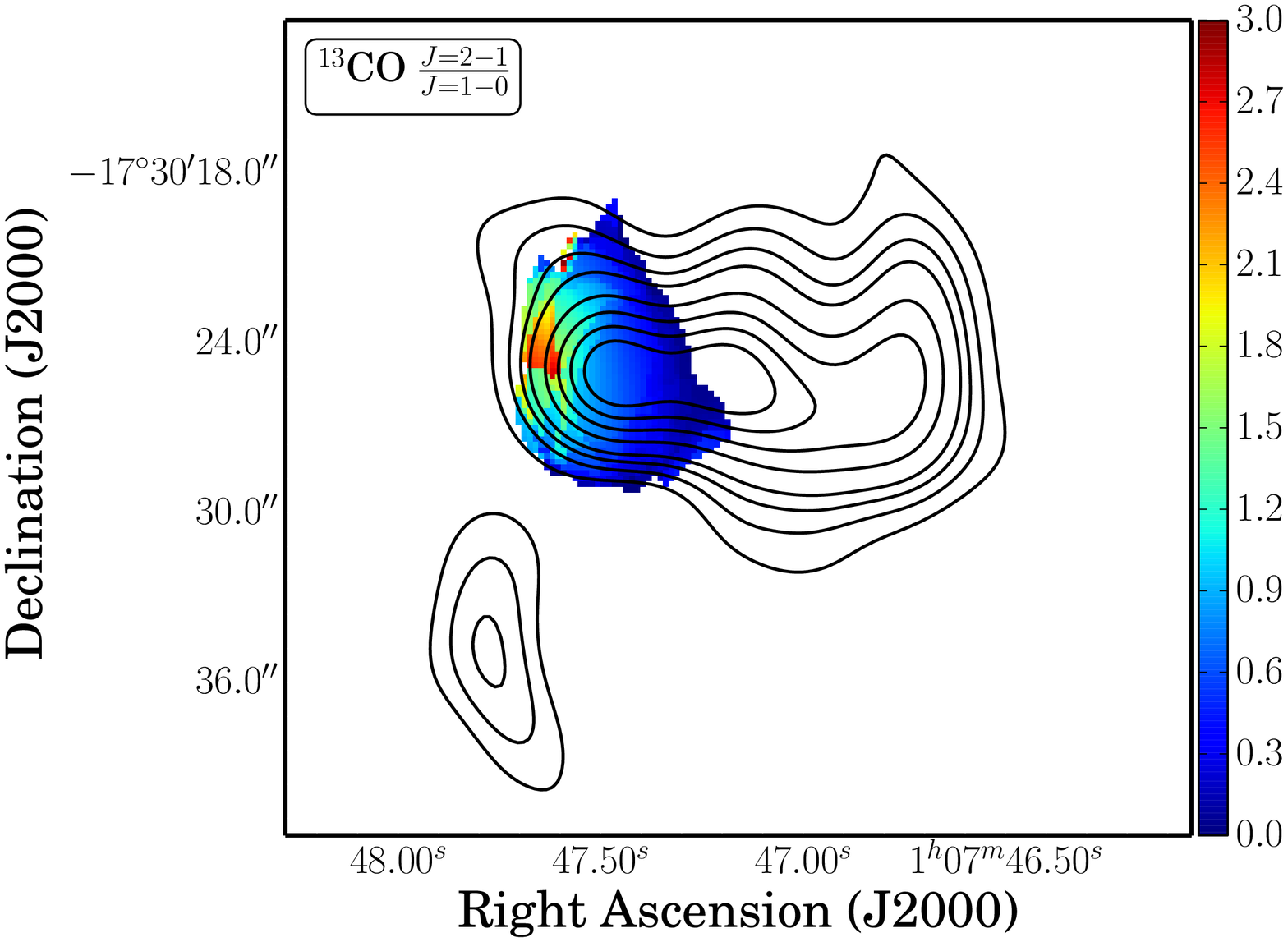} & \includegraphics[scale=0.35]{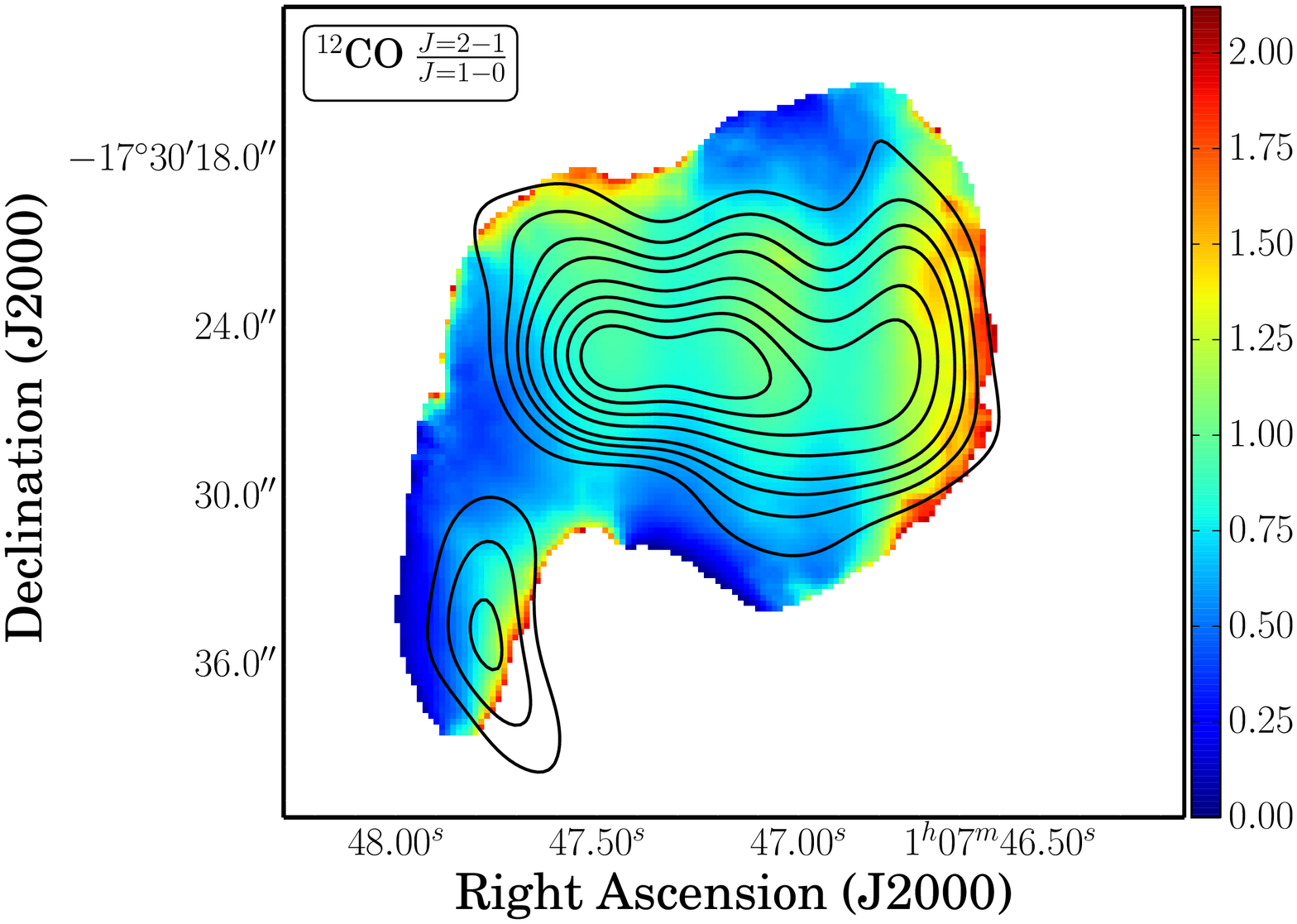} \\
\end{array}$
\caption[Line ratio maps]{Line ratio maps: (\textit{Top Left}) $^{13}$CO $\frac{J=2-1}{J=1-0}$ ; (\textit{Top Right}): $\frac{\mathrm{^{12}CO}}{\mathrm{^{13}CO}}$ $J$=2-1; (\textit{Bottom Left}) $\frac{\mathrm{^{12}CO}}{\mathrm{^{13}CO}}$ $J$=1-0; (\textit{Bottom Right}) $^{12}$CO $\frac{J=2-1}{J=1-0}$. Only emission that is $>$ 3$\sigma$ in both maps is included. Contours of \cotwo\ are shown to aid comparison between the maps. }
\label{RatioMaps}
\end{figure}

\begin{figure}[h] 
\centering
$\begin{array}{c@{\hspace{0.5in}}c}

\includegraphics[scale=0.35]{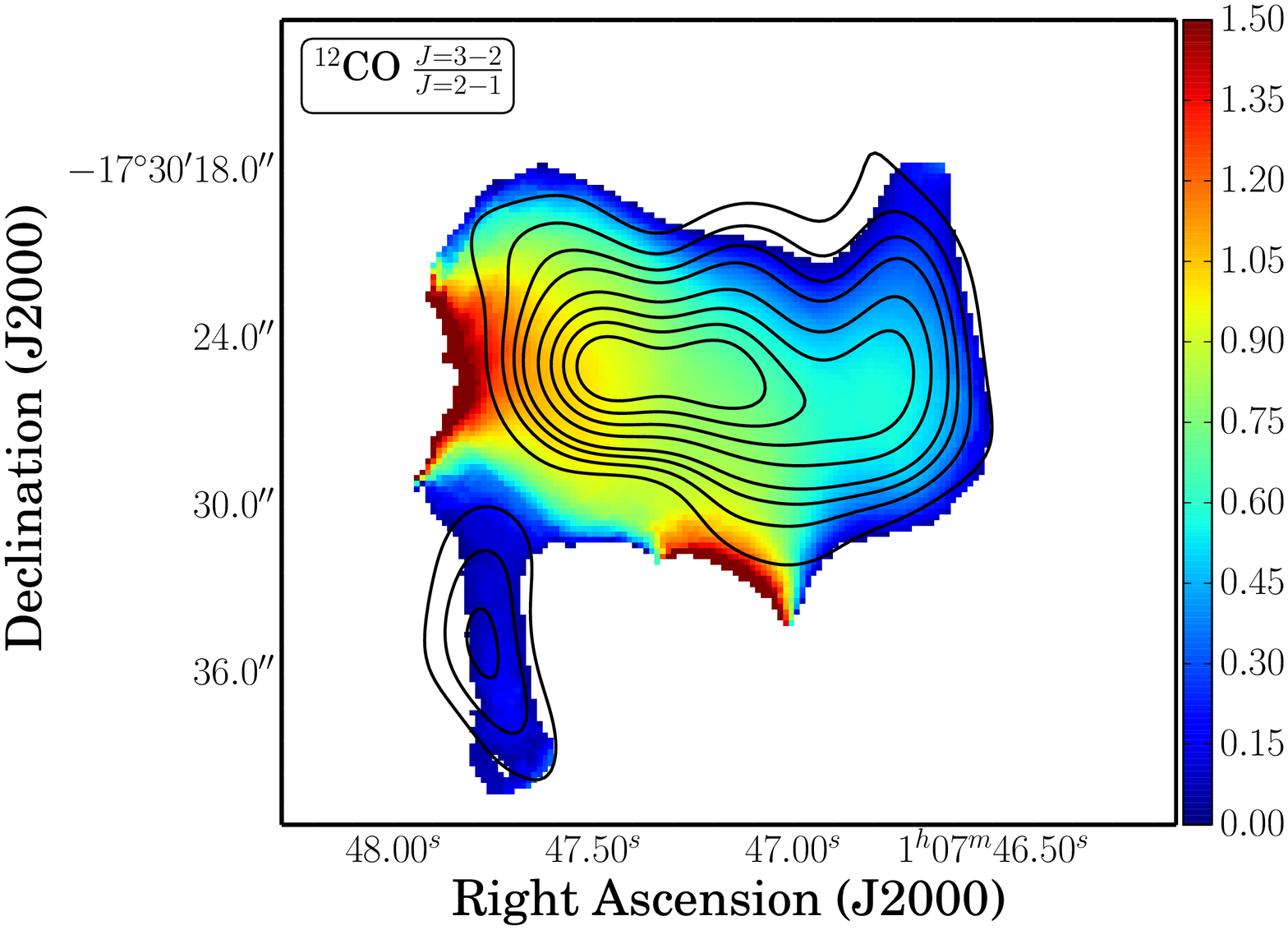} & \includegraphics[scale=0.35]{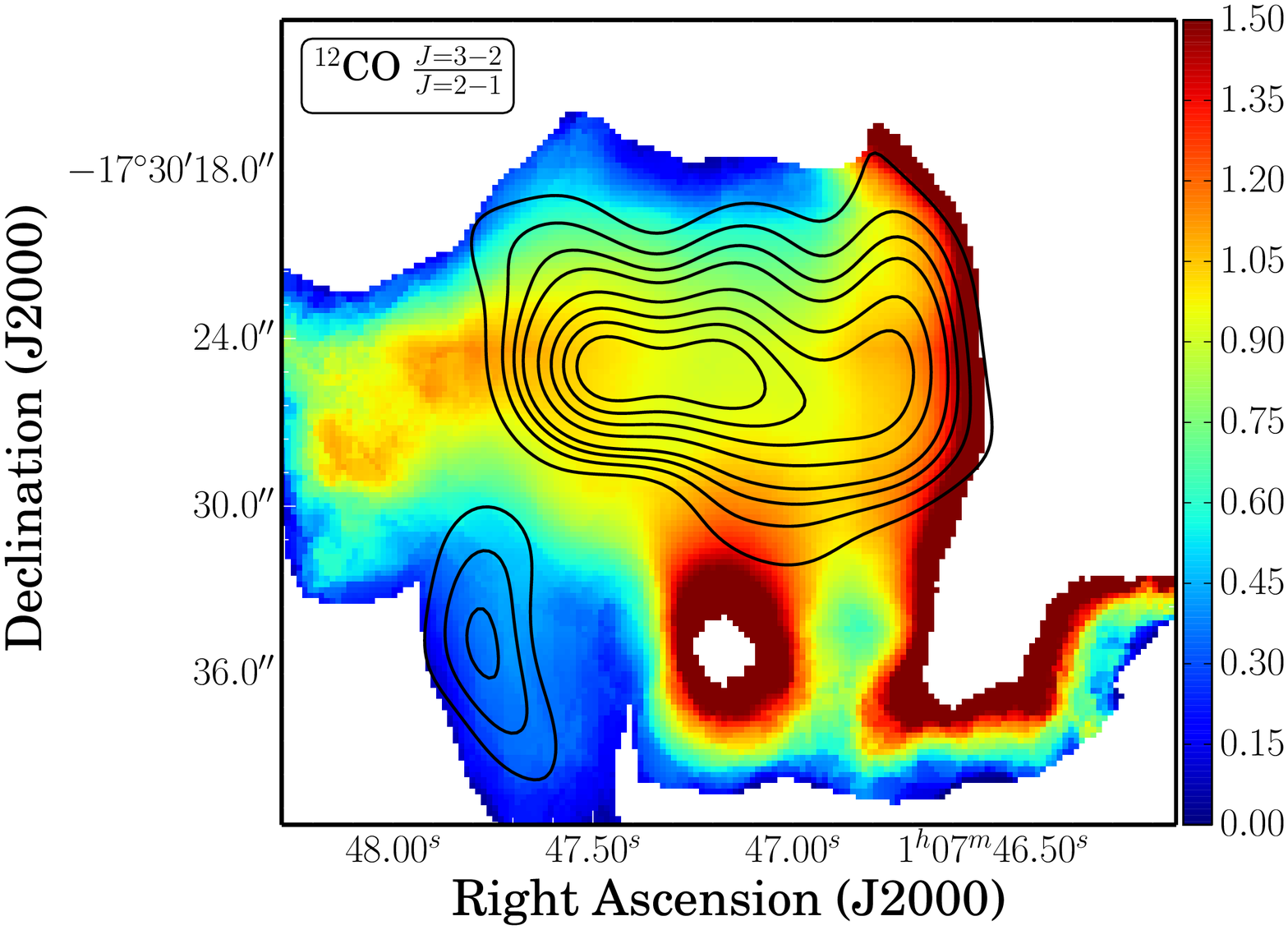} \\ 
 \end{array}$
\includegraphics[scale=0.35]{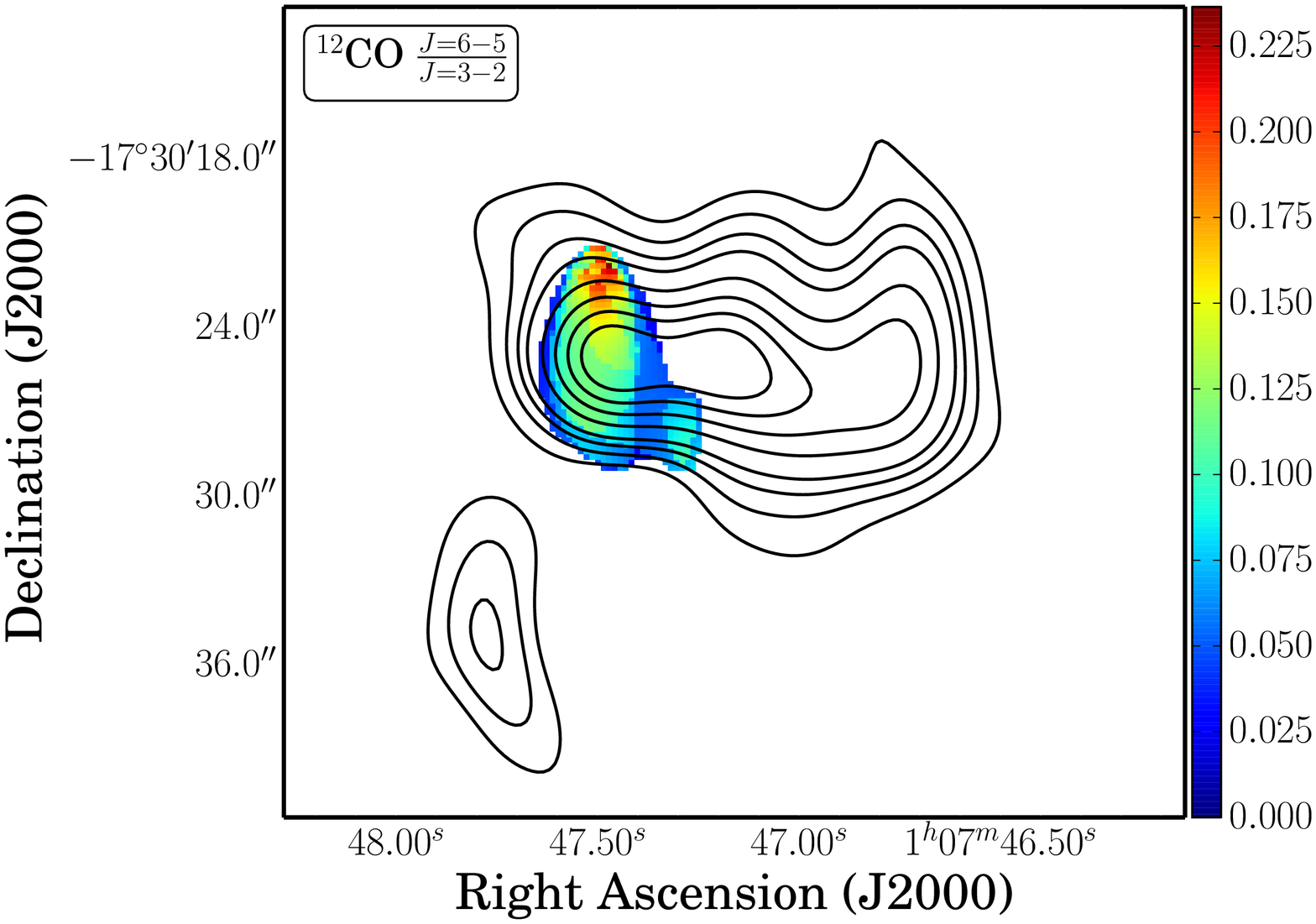} 
\caption[Line ratio maps]{Line ratio maps: (\textit{Top Left}) SMA only $^{12}$CO $\frac{J=3-2}{J=2-1}$; (\textit{Top Right}): SMA + JCMT  $^{12}$CO $\frac{J=3-2}{J=2-1}$; (\textit{Bottom}):  $^{12}$CO $\frac{J=6-5}{J=3-2}$. Contours and emission cutoffs as in Figure 3. }
\label{RatioMaps}
\end{figure}

\begin{figure}[h] 
\centering
\includegraphics[scale=0.8]{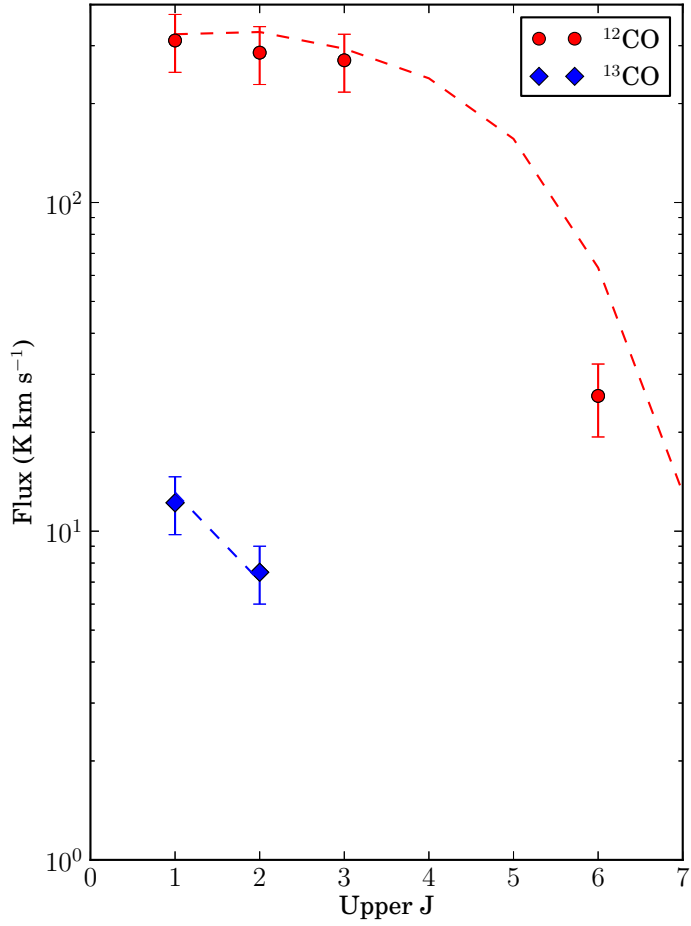} 
\caption[]{$^{12}$CO (\textit{red circles}) and $^{13}$CO (\textit{blue diamonds}) SLEDs of VV 114E using Grid 2 solutions. The dashed lines are the fitted SLEDs from the radiative transfer analysis.}
\label{cosled}
\end{figure}

\begin{figure}[h] 
\centering
$\begin{array}{cc}
\includegraphics[scale=0.45]{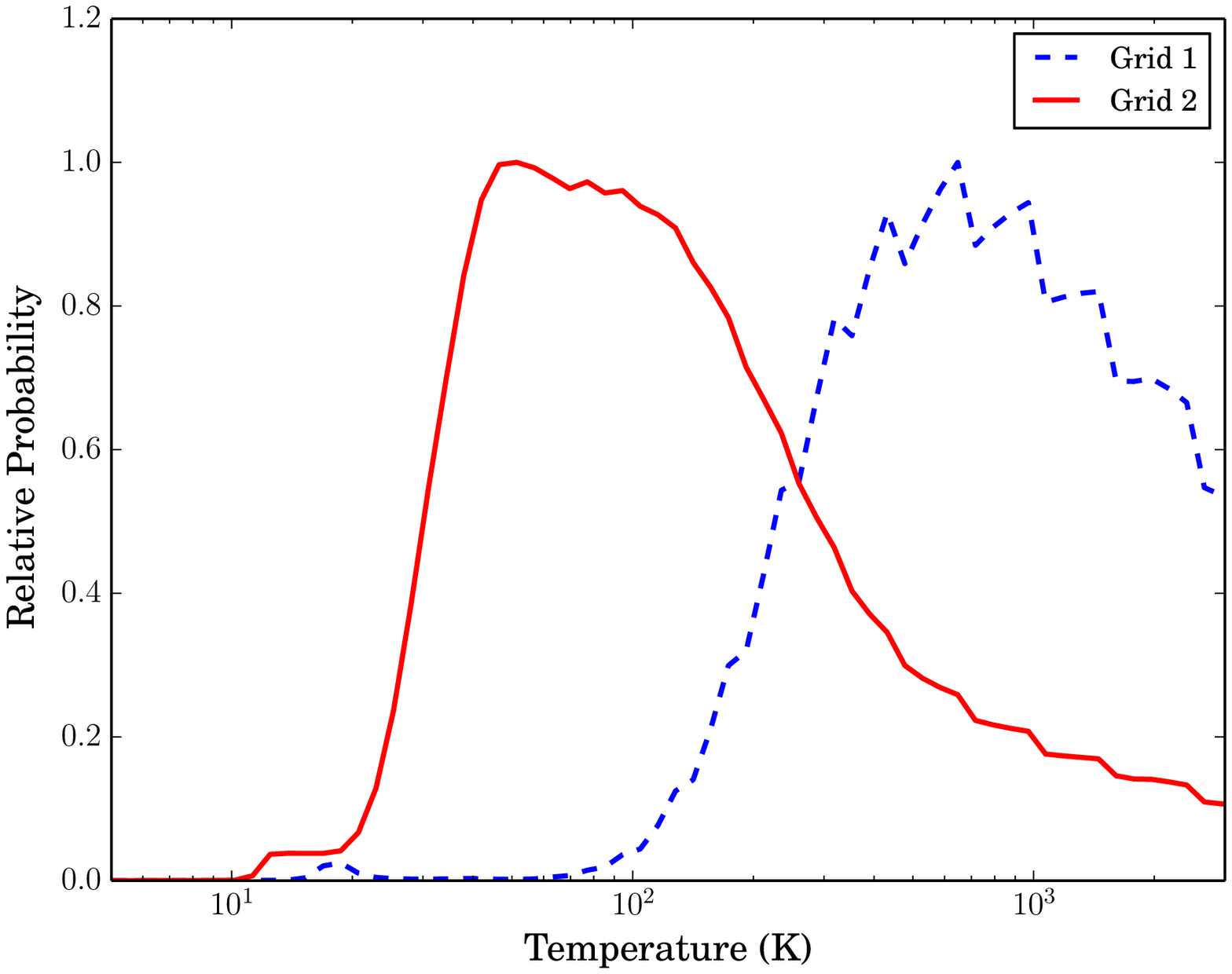} & \includegraphics[scale=0.45]{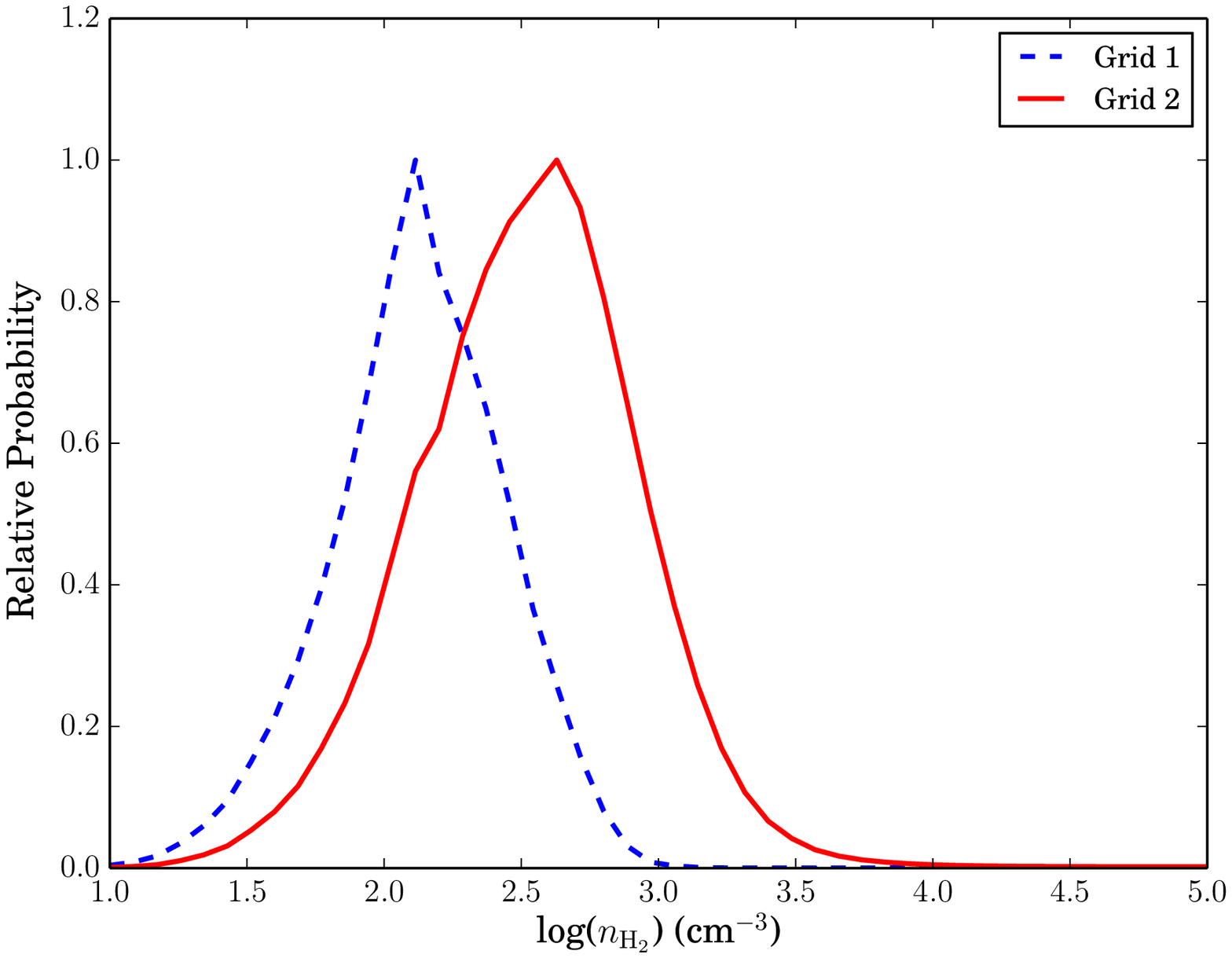} \\ 
\end{array}$
\includegraphics[scale=0.45]{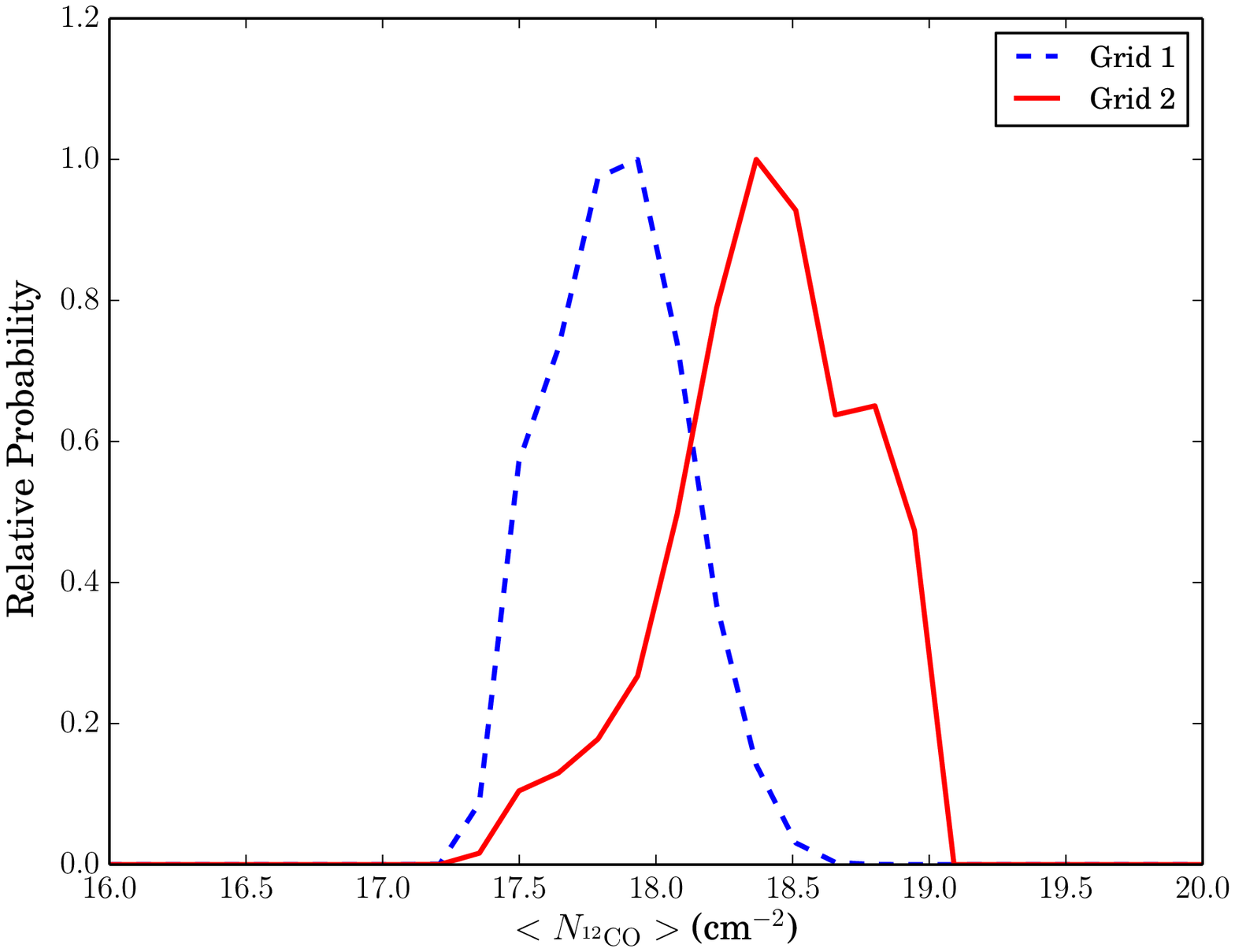}\\
\caption[distrib]{One-dimensional probability distributions for \tkin\ (top), \nhtwo\ (middle) and $<$\nco$>$ (bottom). Both Grid 1 and Grid 2 solutions are shown to display the difference in the solutions with differing \xco\ values. }
\label{}
\end{figure}

\begin{figure}[h] 
\centering
\includegraphics[scale=0.8]{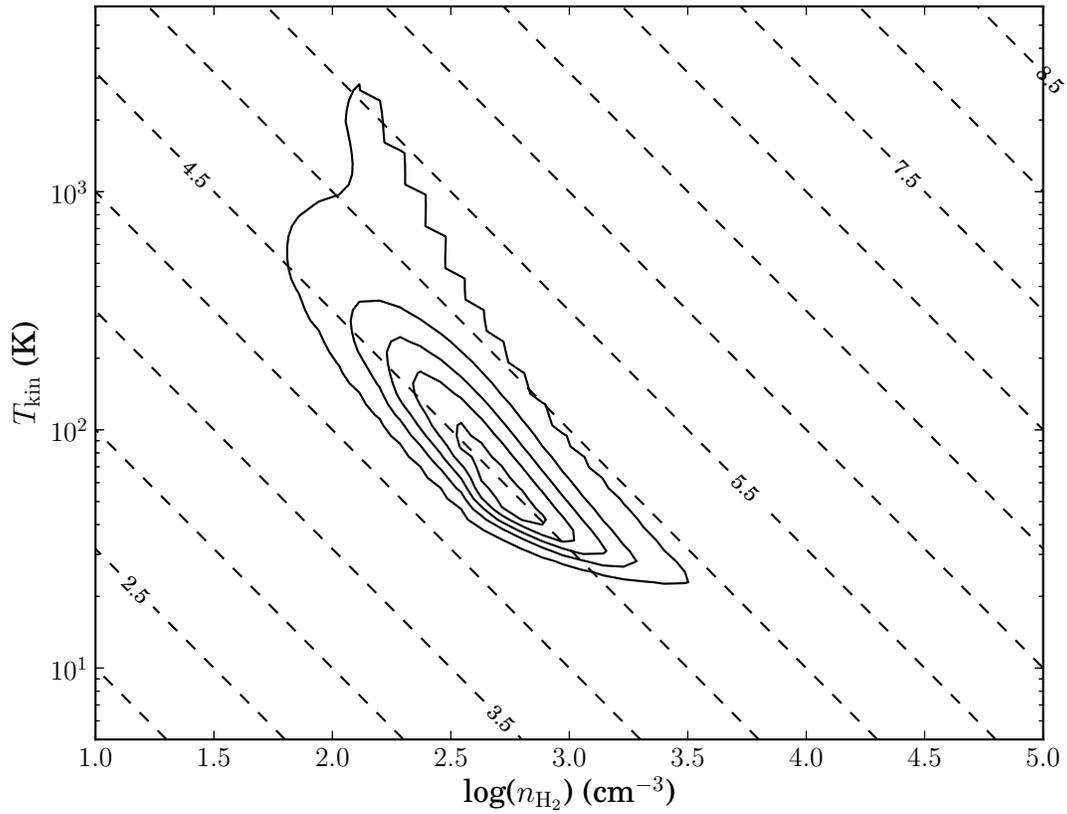} 
\caption[contours]{Two-dimensional distribution for \tkin\ vs. \nhtwo\ using Grid 2 solutions. Diagonal lines indicate pressure in log (K cm$^{-3}$). Contour levels are 10$\%$, 30$\%$, 50$\%$, 70$\%$ and 90$\%$ of the maximum likelihood.}
\label{}
\end{figure}


\begin{deluxetable}{cccccc} 
\tablecolumns{6}
\tablewidth{0pt}
\tablecaption{Data For VV 114}
\label{smasum}
\tablehead{\colhead{Transition Line} & \colhead{Interferometric Flux\tablenotemark{a}} & \colhead{JCMT Flux\tablenotemark{a}} &\colhead{JCMT + SMA Flux\tablenotemark{a}}& \colhead{Beam} & \colhead{rms}\\ \colhead{} & \colhead{(Jy km s$^{-1}$)}& \colhead{(Jy km s$^{-1}$)}& \colhead{(Jy km s$^{-1}$)} & \colhead{($\arcsec$)} & \colhead{(mJy/beam)\tablenotemark{b}} }
\startdata
\coone\tablenotemark{c}	&360 $\pm$ 20		&-&-&6.5 $\times$ 3.7 &25\\
\cotwo				&1280 $\pm$ 50	&2370 $\pm$ 60&2350 $\pm$ 70&4.0 $\times$ 3.0 &13\\
\cothree				&1890 $\pm$ 60	&4800 $\pm$ 200&4800 $\pm$ 60&2.8 $\times$ 2.0 &25\\
\cosix				&150 $\pm$ 40	&-&-&3.1 $\times$  2.2 &270\\
\tcoone\tablenotemark{d}&4.0 $\pm$ 0.3	&-&-&3.7 $\times$ 1.4 &2\\
\tcotwo				&12.0 $\pm$ 2.0	&-	&-&4.6 $\times$ 3.4 &10\\
\enddata
\tablenotetext{a}{measurement uncertainty only; calibration uncertainty is 20$\%$ (Paper I).  }
\tablenotetext{b}{rms values for a 20 km s$^{-1}$ channel width for all lines except \tcoone\ and \tcotwo\ which uses 40 km s$^{-1}$ and 50 km s$^{-1}$ channel widths, respectively.}
\tablenotetext{c}{OVRO map published in Yun et al. (1994)}
\tablenotetext{d}{ALMA Cycle 0 map}

\end{deluxetable}
\begin{deluxetable}{lccccccc} 
\tablecolumns{8}
\rotate
\tablewidth{0pt}
\tablecaption{CO Line Ratios of VV 114\tablenotemark{a}}
\label{line}
\tablehead{\colhead{Region} & \colhead{$^{12}$CO $\frac{J=6-5}{J=3-2}$} &  \colhead{$^{12}$CO $\frac{J=3-2}{J=2-1}$\tablenotemark{b}} &  \colhead{$^{12}$CO $\frac{J=3-2}{J=2-1}$\tablenotemark{c}} &\colhead{$^{12}$CO $\frac{J=2-1}{J=1-0}$} &\colhead{$^{13}$CO $\frac{J=2-1}{J=1-0}$} & \colhead{$\frac{\mathrm{^{12}CO}}{\mathrm{^{13}CO}}$ J=1-0} & \colhead{$\frac{\mathrm{^{12}CO}}{\mathrm{^{13}CO}}$ J=2-1} }
\startdata
VV 114E	&	0.13 $\pm$ 0.05	& 	0.99 $\pm$ 0.28	&0.95 $\pm$ 0.27 & 0.93 $\pm$ 0.26 & 0.63 $\pm$ 0.18 & 25.6 $\pm$ 7.2& 37.7 $\pm$ 5.3\\
VV 114C&	$<$ 0.03\tablenotemark{d} &0.92$\pm$0.26 &0.74$\pm$0.21 &0.89$\pm$0.25 & -- &  $>$ 57\tablenotemark{e} &$>$	77\tablenotemark{e}	\\
VV 114W	&	$<$ 0.04\tablenotemark{d} &1.01$\pm$0.28 &0.57$\pm$0.16 &0.91$\pm$0.26 & --& $>$ 42\tablenotemark{e} &$>$ 60\tablenotemark{e}	\\
\enddata
\tablenotetext{a}{All ratios are in terms of integrated brightness temperatures at the peak position of \cothree. Uncertainties include calibration and measurement uncertainties. All maps were degraded to a resolution of 6.5$\arcsec$ $\times$ 3.7$\arcsec$.}
\tablenotetext{b}{Using short spacing corrected maps.}
\tablenotetext{c}{Using SMA only maps.}
\tablenotetext{d}{Ratios are upper limits using the 3$\sigma$ noise level for \cosix.}
\tablenotetext{e}{Ratios are lower limits using 3$\sigma$ noise level for \tcoone\ and \tcotwo.}
\end{deluxetable}

\begin{deluxetable}{ccccc} 
\tablecolumns{5}
\tablewidth{0pt}
\tablecaption{Source Size and Dynamical Mass}
\label{linewidth}
\tablehead{\colhead{Region} & \colhead{Velocity FWHM\tablenotemark{a}}&  \multicolumn{2}{c}{Deconvolved Source Diameter\tablenotemark{b}} & \colhead{$M_{\rm{dyn}}$} \\ \colhead{} & \colhead{(km s$^{-1}$)}&  \colhead{(arcsec)} & \colhead{(kpc)} & \colhead{(10$^{9}$ $M_{\odot}$)} }
\startdata
VV 114E	&	150	&	(4.1 $\pm$ 0.1) x (1.8 $\pm$ 0.2)	&	(1.71 $\pm$ 0.04) x (0.75 $\pm$ 0.08)	& 	2.4 $\pm$ 0.3\\
VV 114W	&	200	&	(4.4 $\pm$ 0.3) x (2.3 $\pm$ 0.4)	&	(1.8 $\pm$ 0.1) x (1.0 $\pm$ 0.2)	&	5.3 $\pm$ 0.4 \\
VV 114C&	155/95\tablenotemark{c}	&	(7.3 $\pm$ 0.2) x (1.8 $\pm$ 0.3)	&	(3.04 $\pm$ 0.08) x (0.8 $\pm$ 0.1)	&	--\\
\enddata
\tablenotetext{a}{Uncertainty in velocity FWHM is 10 km s$^{-1}$}
\tablenotetext{b}{Beam size is 2.8$\arcsec$ $\times$ 2.0$\arcsec$}
\tablenotetext{c}{The central region spectrum is fitted with two Gaussian peaks.}
\end{deluxetable}

\begin{deluxetable}{ccccc} 
\tablecolumns{4}
\tablewidth{0pt}
\tablecaption{Modeling Results\tablenotemark{a}}
\label{results}
\tablehead{ \colhead{} & \colhead{1DMax} & \colhead{4DMax} & \colhead{1$\sigma$ range}  }
\startdata
\tkin\ (K) &	51&		38&		42 -- 459  \\
\nhtwo\ (cm$^{-3}$) &	10$^{2.63}$ &		10$^{2.89}$&		10$^{2.07}$ -- 10$^{2.88}$  \\
\nco\ (cm$^{-2}$) &		10$^{19.58}$ &	10$^{20.03}$& 	10$^{18.87}$ -- 10$^{19.80}$   \\
\ff\ & 				10$^{-1.14}$ &	10$^{-1.14}$ &	10$^{-1.26}$ -- 10$^{-0.87}$   \\
$P$ (K cm$^{-3}$) &	10$^{4.58}$  &		10$^{4.58}$  & 	10$^{4.33}$ -- 10$^{4.88}$   \\
$<$\nco$>$ (cm$^{-2}$) & 10$^{18.37}$ & 	10$^{18.37}$ &	10$^{17.98}$ -- 10$^{18.71}$   \\
Mass (\msol) &			10$^{8.79}$& 	10$^{8.79}$&		10$^{8.40}$ -- 10$^{9.13}$ &   \\
\hline
\hline
\xco & 229 & 229 & 112 -- 603 \\
\nthreeco (cm$^{-2}$) & 10$^{17.13}$ & 10$^{17.13}$ & 10$^{16.75}$ -- 10$^{17.18}$ \\
\enddata
\tablenotetext{a}{ These are the modeling results for Grid 2.}

\end{deluxetable}

\end{document}